\documentclass[final,authoryear,5p,times,twocolumn]{elsarticle}

 \usepackage{graphics}

\usepackage{amssymb}

 \usepackage{lineno}
 \usepackage{rotating}
 \usepackage{multirow}
\usepackage{lscape}
\usepackage{color} 

\journal{Icarus}

\begin{document}

\begin{frontmatter}


 \title{Title\tnoteref{label1}}
\tnotetext[label1]{Universiteit Utrecht, Faculty of Geosciences,
Budapestlaan 4, 3584 CD Utrecht, The Netherlands}
 \title{Title\tnoteref{label1}}
\tnotetext[label2]{Jacobs University Bremen, Campus Ring 1, 28759 Bremen, Germany}
 \author{Manuel Roda\corref{cor1}\fnref{label1}}
  \ead{M.Roda@uu.nl}
 \author[label1]{Maarten G. Kleinhans}
  \ead{M.G.Kleinhans@uu.nl}
 \author[label1]{Tanja E. Zegers}
 \ead{T.E.Zegers@uu.nl}
 \author[label2]{Jelmer H.P. Oosthoek}
 \ead{j.oosthoek@jacobs-university.de}
 \cortext[cor1]{Corresponding author}

\title{Catastrophic ice lake collapse in Aram Chaos, Mars}




\begin{abstract}
Hesperian chaotic terrains have been recognized as the source of outflow channels formed by catastrophic outflows. Four main scenarios have been proposed for the formation of chaotic terrains that involve different amounts of water and single or multiple outflow events. Here, we test these scenarios with morphological and structural analyses of imagery and elevation data for Aram Chaos in conjunction with numerical modeling of the morphological evolution of the catastrophic carving of the outflow valley. The morphological and geological analyses of Aram Chaos suggest large-scale collapse and subsidence (1500 m) of the entire area, which is consistent with a massive expulsion of liquid water from the subsurface in one single event. The combined observations suggest a complex process starting with the outflow of water from two small channels, followed by continuous groundwater sapping and headward erosion and ending with a catastrophic lake rim collapse and carving of the Aram Valley, which is synchronous with the 2.5 Ga stage of the Ares Vallis formation. The water volume and formative time scale required to carve the Aram channels indicate that a single, rapid (maximum tens of days) and catastrophic  (flood volume of 9.3$\cdot$10$^4$ km$^3$) event carved the outflow channel. 
We conclude that a sub-ice lake collapse model can best explain the features of the Aram Chaos – Valley system as well as the time scale required for its formation.
\end{abstract}


\end{frontmatter}


\section{Introduction}
\label{Introduction}
Martian chaotic terrains are deeply collapsed areas ($>$ 1 km deep) that stretch for up to hundreds of kilometers and show a bumpy floor characterized by an irregular pattern of fractures and tilted blocks of different sizes (from meter to kilometer scale). Chaotic terrains predominantly occur along the dichotomy boundary between the southern highlands and northern lowlands \citep{Sharp1973,Chapman2002,Rodriguez2005a,Glotch2005,Meresse2008,Warner2011}.

Outflow channels represent the largest systems carved by liquid water on Mars. They are thousands of kilometers long, more than a kilometer deep \citep{Baker2001}and show attributes such as grooves, terraces, teardrop islands, streamlined terraces and high width-to-depth ratios that are consistent with the erosive origin of the channels \citep[e.g.,][]{Nelson1999,Baker2001,Coleman2005,Pacifici2009,Warner2010a}.

In many cases, chaotic terrains represent the source area of Hesperian (approximately 3.7--3.3 Ga) outflow channels \citep{Nelson1999,Tanaka2003} and several authors \citep[e.g.,][]{Carr1979,Carr1996a,Baker2001} argue that those chaotic terrains were formed by a rapid discharge of water from the subsurface, resulting in collapsed and fractured areas and massive flows carving the large outflow channels. However, the actual evolutionary process leading to chaotic terrain formation and collection and the discharge of catastrophic volumes of water ($\ge$10$^5$ km$^3$) has remained controversial.

Several evolutionary processes have been proposed that can be grouped in four different scenarios (Fig. 1). In the first hypothesis (Fig. 1a), fracturing in the bedrock led the water expulsion to the surface. The water was generated by partial melting of the cryosphere after magmatic intrusions that increased the subsurface temperature \citep[e.g.,][]{Sharp1973,Chapman2002,Ogawa2003,Rodriguez2005a,Leask2006b,Meresse2008}.

\begin{figure*}[ht!]
\centerline{\includegraphics[width=.75\textwidth]{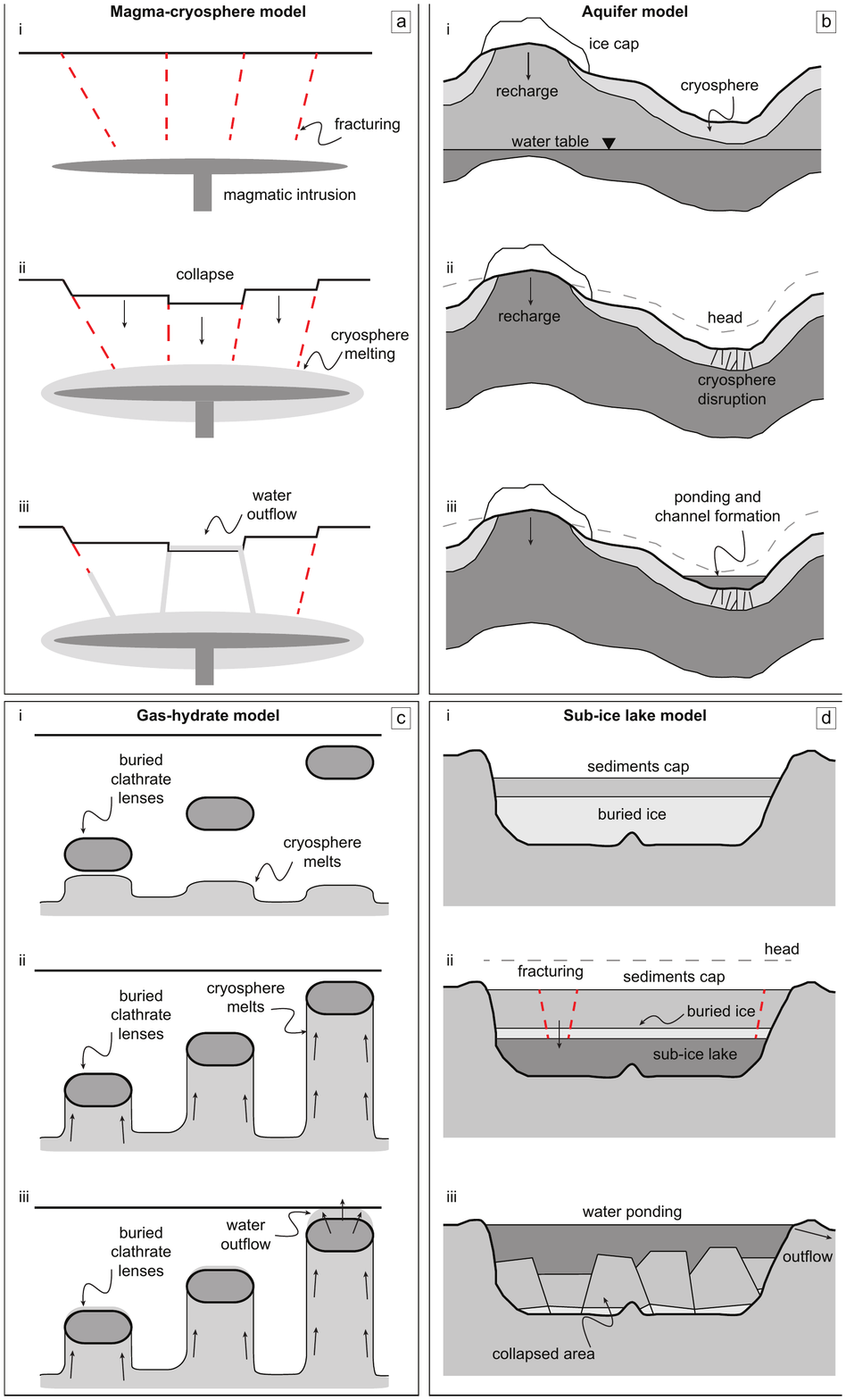}}
\caption{Four different scenarios for the origin of chaotic terrains. a) Magma-cryosphere model \citep[redrawn after][]{Meresse2008}; b) Aquifer model \citep[redrawn after][]{Harrison2009}; c) Gas-hydrate model \citep[redrawn after][]{Kargel2007}; d) sub-ice lake model \citep[redrawn after][]{Zegers2010}.}
\label{map}
\end{figure*}

In a second group of mechanisms (Fig. 1b), the release of water from the cryosphere was the result of an increase in pressure of a global pressurized sub-cryospheric aquifer \citep[e.g.,][]{Carr1979,Clifford1993,Hanna2007a,Harrison2009}. The discharge from the pore space relied on the flow of water through a permeable subsurface layer, where water that was discharged was replaced by recharge. In the majority of martian aquifer models, water is generally assumed to have recharged from a great distance \citep[$\ge$2000 km,][]{Clifford1993,Harrison2008}.
Numerical flow models \citep{Hanna2007a,Harrison2008} indicate that the large total volume of liquid water (10$^5$--10$^6$ km$^3$) and high discharge rates (10$^6$--10$^9$ m$^3$/s) required to form the morphology of the outflow channels were not achievable by a single discharge from a porous medium. Because of the flow volume discharge from aquifers, a scenario was proposed wherein a large number of small flooding events were followed by a sudden release of previously ponded water \citep[$\ge$600,][]{Harrison2008}.

Another hypothesis (Fig. 1c) suggests that dewatering of fluids from gas and/or salt hydrate buried deposits \citep{Max2001,Montgomery2005} or the hydrologic processes triggered by clathrates \citep{Kargel2007} could be responsible for the water outflow.

Finally, \citet{Zegers2010} propose that chaotic terrains developed by the catastrophic collapse of sediments induced as a consequence of the melting of buried ice sheets (Fig. 1d). Thermal modeling results show that even under very low crustal heat flux conditions, ice sheets will melt if buried under thick sediments (up to 2 km) because of the difference in thermal conductivity between the basin fill and surrounding crust. When the buried sub-ice lake reaches a critical thickness, the overburden collapses and subsides, resulting in a massive expulsion of water to the surface.

To distinguish between those four scenarios and test the validity of the evolution models for chaotic terrains and their outflow channels, the amount and timing of the water release, the amount of subsidence, and the fracture distributions are fundamental variables. To constrain those variables for a single chaotic terrain and its outflow channel, we analyzed the morphological and geological features characterizing Aram Chaos and its valley. Furthermore we estimated the flow volume and formative time scale required to carve the Aram channel. Finally we discuss the four scenarios proposed for chaotic terrain formation in view of these results, propose the most likely scenario for the evolution of Aram Chaos and Aram channel and discuss the significance of our findings for other chaotic terrains.

\section{Aram Chaos morphology and fracturing}
In this section, we present the geological and morphological observations related to Aram Chaos in conjunction with a structural analysis performed on the fractures to determine the sequence of depositional and fracturing events during its formation.

\subsection{Geology and pre-faulting morphology}
Aram Chaos is situated in a circular basin with a diameter of 280 km centered at 2.5$^{\circ}$N and 338.5$^{\circ}$E (Fig. 2) suggesting that it is developed in a large impact crater \citep{Schultz1982,Glotch2005}. It is connected to Ares Vallis by Aram Valley, a 15 km wide and 2.5 km deep outflow channel \citep{Glotch2005,Masse2008,Oosthoek2007a}.
The formation time of the original impact crater is not constrained; however, given the size of the crater, the impact likely occurred in the Noachian \citep[$\ge$ 3.7 Ga,][]{Zegers2010}.It has been suggested that the outflow event of Aram Valley was synchronous with the final erosive event of Ares Vallis \citep[approximately 2.5 Ga,][]{Warner2009}.

\begin{figure}[ht!]
\centerline{\includegraphics[width=\columnwidth]{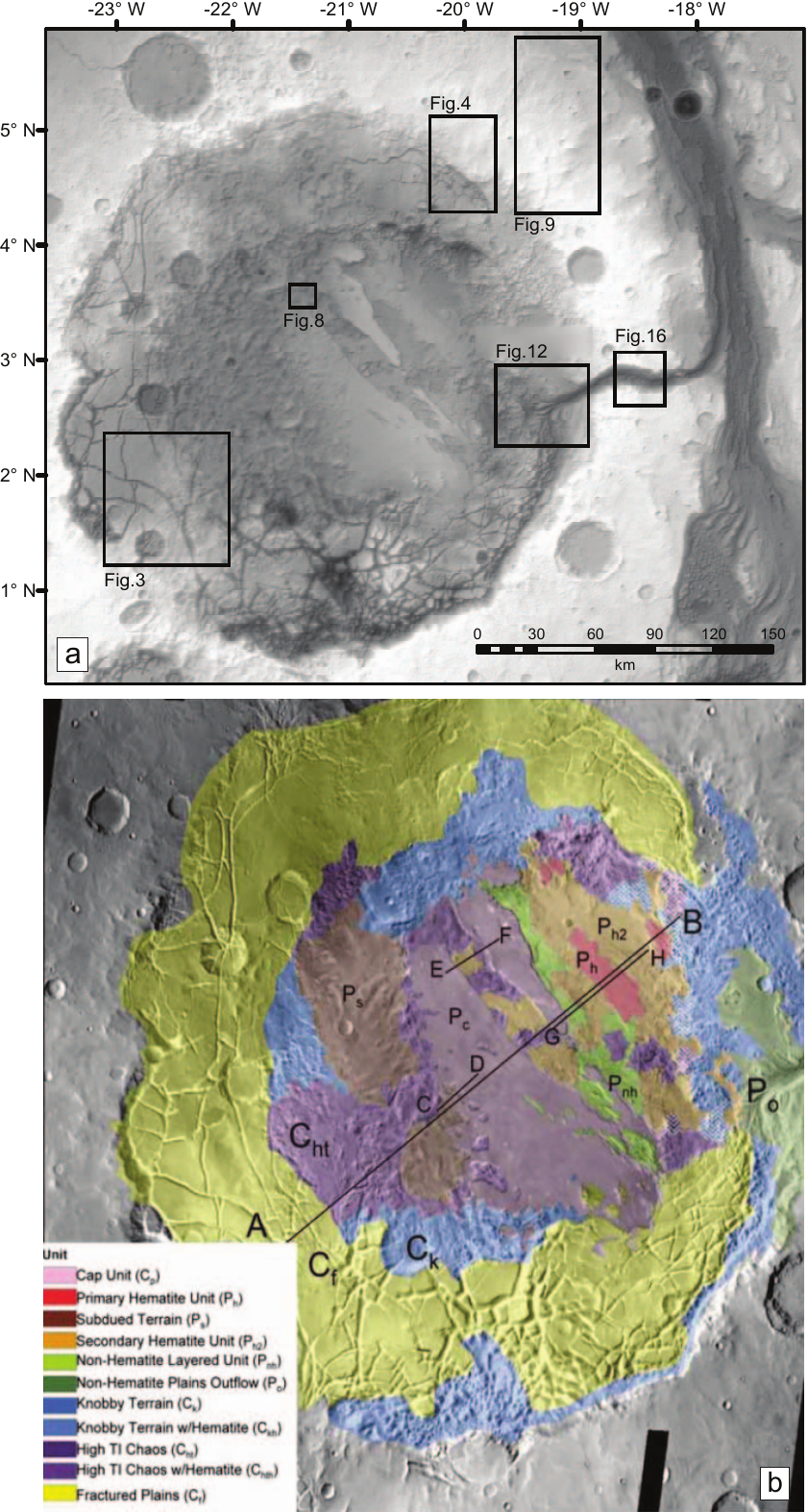}}
\caption{a) View of Aram Chaos showing the typical fractured and tilted chaos blocks, and associated outflow channel (Aram Valley). The black squares are referred to the locations of detailed figures (HRSC mosaic). b) Geological map of Aram Chaos after \citet{Glotch2005}. This material is reproduced with permission of John Wiley \& Sons, Inc.}
\label{map}
\end{figure}

\citet{Glotch2005} divide the Aram chaotic terrain into three possibly lateral main units. The Fractured Plains unit is the largest and almost completely surrounds the crater rim. It consists of up to tens of kilometer-sized slumped blocks forming a curvilinear fracture pattern. The Knobby Terrain unit is the second unit and occurs in the central part of Aram Chaos and at certain locations surrounding the crater rim. It consists of km-scale irregular blocks (knobs). The High Thermal Inertia Chaotic Terrain unit occurs in the central part of Aram Chaos and underlies outcrops of layered material. Those two \citep{Masse2008} or three \citep{Lichtenberg2010} layered subunits (500-m thick) are composed by mono- and polyhydrated minerals such as sulfates and ferric-oxides.

Observations in 3D, particularly from HRSC stereo data (12 m/pixel), provide evidence that the current Aram Chaos terrain was originally a large impact basin that was almost entirely filled by sediments before fracturing and subsidence occurred and resulted in the present-day morphology. Figure 3 shows a part of the rim where evidence for this sequence of events is clearly visible from a view based on HRSC data. In the foreground of figure 3a, a location is visible where the original boundary between the crater rim highlands and crater fill is not disrupted by fractures. The transition from crater rim to crater fill in this location is similar to many filled craters in the highlands of Mars \citep{Warner2010b}, which include a crater rim terrain showing irregular erosional features with an onlap transition to smooth crater fill material. There is a gradual slope from the crater rim into the crater fill of almost 500 m (Fig. 3b, white arrow), which suggests that the crater was likely filled to a level of 500 m below the rim in this part. In the background of figure 3b, the transition from unfractured highland to chaotic terrain is visible, and the transition is marked by an abrupt escarpment of 1500 m.

Some impact craters along the northern rim (Fig. 4a and b) indicate that there was a continuous surface between the surrounding highland terrain and Aram crater fill material prior to fracturing. We conclude that the pre-faulting geometry of the basin was similar to many other large impact craters on Mars: the crater was almost entirely filled, but the outline of the crater is still visible (Fig. 4). In the case of Aram Chaos, the crater was likely filled to a level 500 m below the rim in the south part and almost completely filled in the northern part, as indicated by the different characters of the transition between highland and crater fill (Figs 4b and 5b). This suggests that the difference in elevation between the northern and southern part was almost 500 m before the collapse, supposing a homogeneous rim degradation. Aram Chaos is in a region thought to have experienced massive erosion and deposition of a sedimentary overburden during the Late Noachian-early Hesperian \citep{Hynek2001}. Based on crater counting, \citet{Warner2010a} suggest that the intense degradation and infill occurred during a short 200 Myr interval in the Late Noachian, from 3.8 Ga to 3.6 Ga. Although sedimentation rates were likely higher in the Noachian and spatially variable \citep{Grotzinger2012}, even very low sedimentation rates of 0.01 mm/yr would have been sufficient to fill a deep (2 km) depression in a 200 Myr time span.

\begin{figure}[ht!]
\centerline{\includegraphics[width=1\columnwidth]{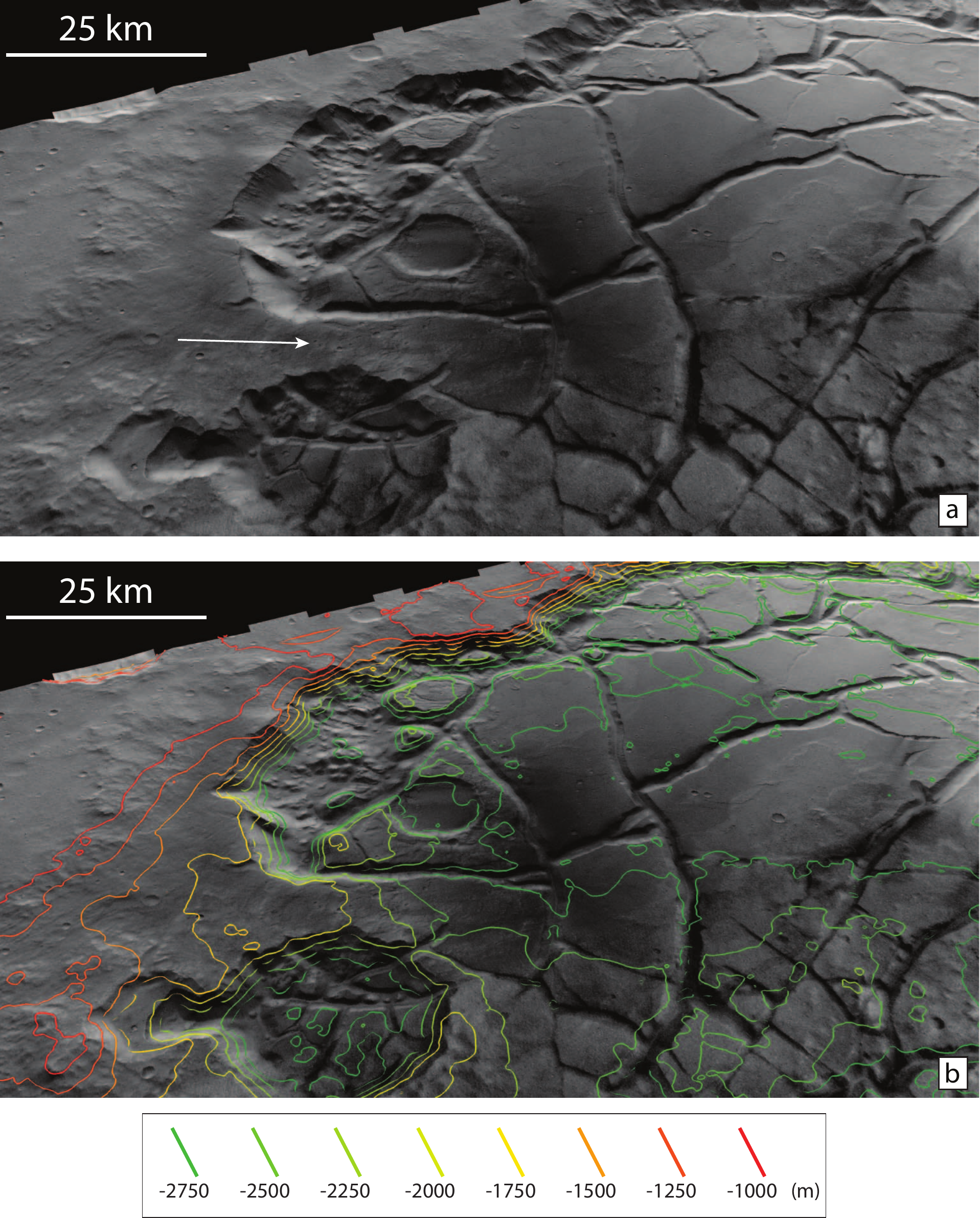}}
\caption{a) Perspective view, looking towards the north-west of the south-western part of the rim of Aram Chaos basin. The white arrow indicates where the original boundary between the crater rim highlands and the crater fill is not disrupted by fractures. There is a gradual slope from the crater rim into the crater fill. Lower image (b) shows the elevation contour lines. In the background the elevation drop between unfractured and fractured terrain is at least 1000 m between. Image based on HRSC mosaic (orbits $H0401\_0001$ and $H1000\_0000$) draped over HRSC-derived Digital Elevation Model.}
\label{topoS}
\end{figure}

The original depth of the Aram Chaos crater was estimated using an empirical relation between crater diameter and pristine depth \citep{Parker2010}:
\begin{equation}
H(m) = 350\cdot D^{0.44}(km)
\end{equation}
For Aram Chaos, with an estimated diameter ($D$) of 280 km, this resulted in an original depth ($H$) of 4.2 km. This suggests that prior to fracturing and subsidence, the original Aram Chaos crater of approximately 4 km depth was filled by sediments to at least 500 m below the rim (south part) or until the rim (north part).
 
\subsection{Fracturing and depositional events}
Aram Chaos shows a fractured and collapsed morphology that resulted in a geometry of faulted blocks that is characteristic of chaotic terrains. The rim faults have experienced most of the displacement related to subsidence and are up to 1500 m on the western rim. Fault displacements between individual fault blocks in the basin are minor compared to the rim fault.

The fracture pattern was analyzed in terms of fracture density and fracture orientation (Fig. 5a) to infer the fracture mechanism. The fracture spacing is highly variable and ranges from tens of kilometers down to locally small values below the resolution limit where no single block is detectable.
No preferred orientation is obvious in the fracture patterns of the fractured units (Fig. 5b). Only a weak alignment of fractures with the rim orientation exists in the outer zone. The fracture pattern includes interfering patches of radial fractures, (Fig. 5a, arrows), which partly originated from loci of very high fracture density. The fracture density increases in the area around the outflow channel (north and south), and a more abrupt escarpment characterizes the southern rim of the Aram crater with respect to the northern part, which presents a more gradual slope of the rim (Fig. 6).

\begin{figure}[ht!]
\centerline{\includegraphics[width=\columnwidth]{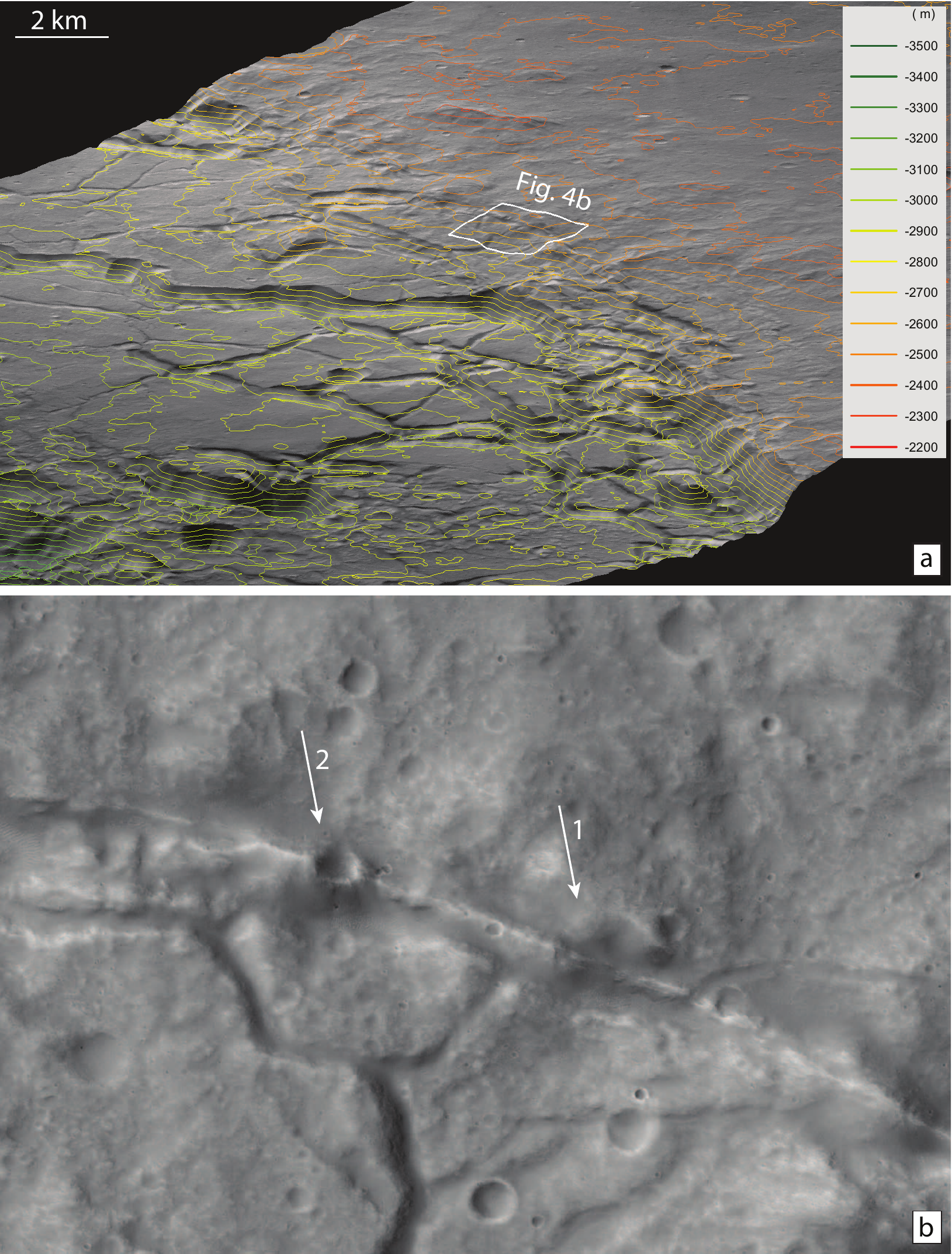}}
\caption{a) Perspective view, looking towards the north-west of the northern part of the rim of Aram Chaos basin. In the center of the image the continuity between the crater filling and the highlands is clearly visible suggesting that the crater was almost entirely filled before fracturing. Image based on HRSC mosaic (orbits $H0967\_0000$) draped over HRSC-derived Digital Elevation Model. b) Detail of the Aram Chaos northern rim: along the boundary two small impact craters illustrate the ancient (crater 1) and present-day (crater 2) continuity between Aram Chaos sedimentary fill and the highland terrains. The rim is clearly identifiable on the floor of crater 1 but it is deleted in crater 2. CTX image P03\_002272\_1829\_XI\_02N020W.}
\label{topoN}
\end{figure}

On the basis of fracture pattern and morphology, \citet{Oosthoek2007a} differentiated between two different fractured units: Highland and Chaos Formation and Lower Aram Chaos Formation (Fig. 7). The latter is further divided into three lateral subunits: fractured, broken and smooth. The fractured subunit is cross-cut by relatively small-scale fractures (1--2 km scale) compared to the fractures of the Highland and Chaos Formation. Some fractures have raised rims and others show small thrusts at the base of the rim (Fig. 8). The broken subunit is highly fractured with approximately 1-km sized irregular mesas. This subunit always occurs at the boundary of the fractured Highland and Chaos Formation. The fractured and broken units are contained by the approximately 150 km diameter inner ring (Figs. 5 and 7) of the Aram Chaos crater. The smooth subunit is non-fractured and may in fact be a relatively thin unit covering the fractured subunits (see Fig. 7).

These fracture-geological unit relationships may either be interpreted as two distinct fracture events or as one continuous fracture event, during which the fractured Aram unit was deposited. The first event is represented by the fracturing and collapse of the Highland and Chaos Formation. The fractured and broken subunit of the Lower Aram Chaos Formation was deposited during (or just after) the collapse of the underlying Highland and Chaos Formation and has subsequently been fractured and broken up. The smooth subunit of the Lower Aram Chaos Formation represents the first depositional event after the collapse covering the fractured subunits. Intermediate and Upper Aram Chaos Formations are not fractured and are deposited on top of the fractured units, closing the depositional sequence.

Geological, morphological and structural analyses on Aram Chaos suggest a large-scale collapse and subsidence of the entire area (1000--1500 m).

\section{Outflow channels morphology}
A large valley (Aram Valley) cuts through the entire eastern rim of Aram Chaos. On the northeastern rim, two minor channels are visible. In this section, we describe the morphological and morphometric features of the outflow channels to determine the chronological sequence of the outflow events.

\subsection{Minor channels} 
Along the northeast rim of Aram Chaos, two small channels are visible (Fig. 9). They have a similar and constant slope (0.019 for the northern and 0.020 for the southern, measured from HRSC dataset) and a general U-shaped profile (Fig. 10). The northern channel is 5 km wide with a width/depth ratio of 50; the southern channel is 4 km wide with a width/depth ratio of 40. Their inlets are located at approximately -1800 m elevation (Fig. 10b). The outlets are truncated by the erosional features of Ares Vallis (Fig. 9) at an elevation between -3500 and -4000 m. These features represent the remnants of the late erosive events in Ares Vallis \citep[2.8--2.5 Ga,][]{Warner2009}. This indicates that the two outflow channels were active before or during the late carving of Ares Vallis. Furthermore, the high channel slopes that are quite similar to the slopes of the pristine Aram Chaos rim, suggest that they could be representatives of the early outflow stages recorded in the Aram Chaos. They may also have resulted from groundwater sapping toward Ares Vallis that occurred along Aram crater walls saturated with fluid. 

Whether these small channels formed as water sloshed over the rim after collapse or as sapping channels in the Aram crater wall saturated with fluid water, water must have been present within Aram Chaos. Assuming an empty crater and water-saturated rims, water flow from the rims toward the crater would be expected because of a higher subsurface gradient along its direction, with a consequent opposite carving direction of minor channels. A rainfall origin for those channels (unlikely during the Late Hesperian, i.e., after the collapse) is inconsistent with the absence of other small channels around the Aram Chaos rim.

\begin{figure*}[ht!]
\centerline{\includegraphics[width=0.8\textwidth]{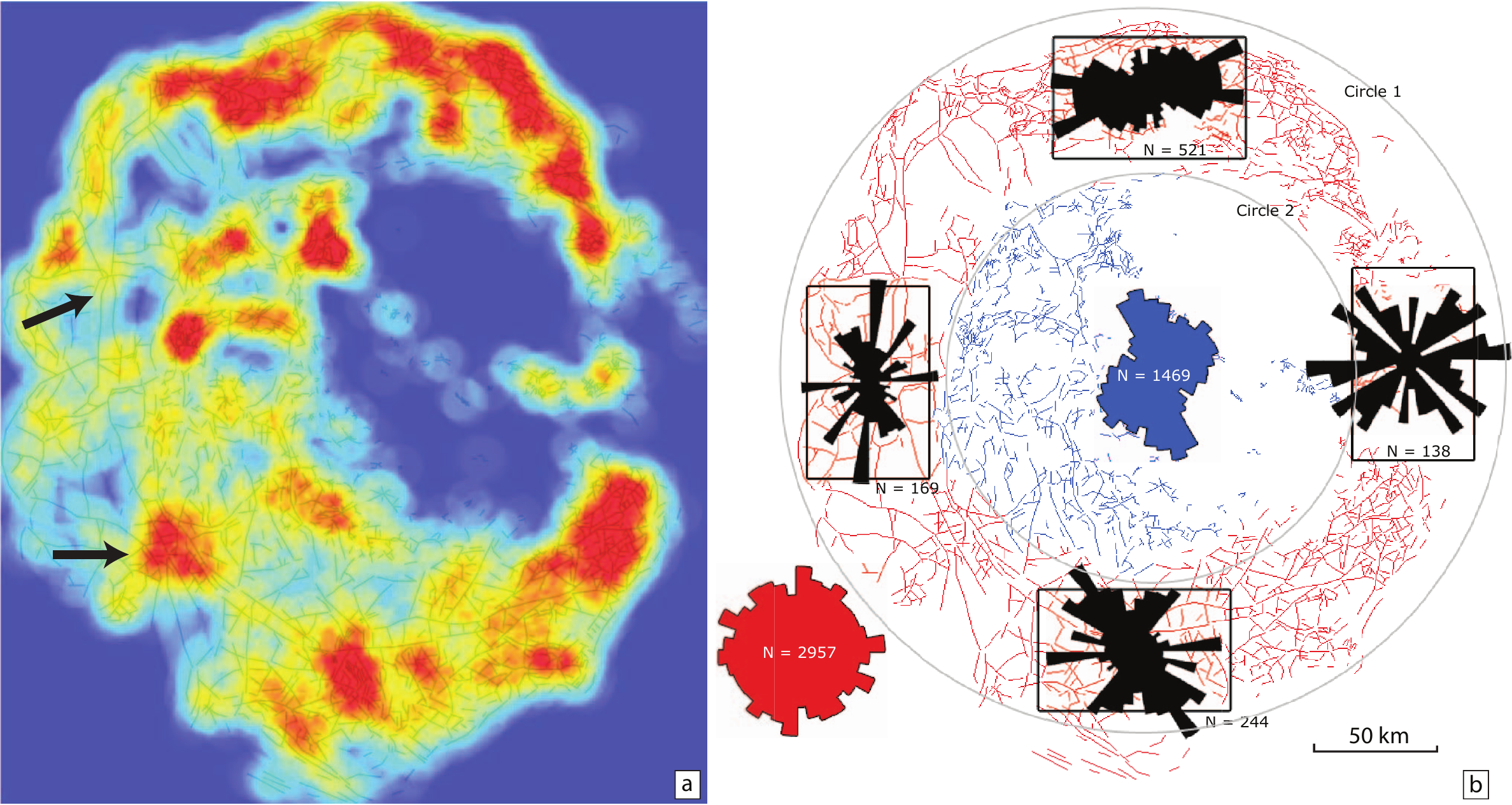}}
\caption{a) Map of Aram chaos showing the asymmetric fracture density (lineament length per km$^2$, dark red fracture density is 0.387 km$^{-1}$). Semi-radial fractures patterns are indicated with arrows. b) Fracture orientation in the Aram Chaos region. In red the fractures formed in the outer ring, predominantly in the Highland Unit, in blue the fractures formed in the center of the chaotic terrain, predominantly in the fractured Aram unit. The length-weighted rose diagrams in black represent the orientations of fractures in the squares indicated in Aram Chaos. The rose diagram in red (blue) represents all fractures measured in the red (blue) area.}
\label{fractures}
\end{figure*}

\subsection{Aram Valley} 
The Aram Valley is a deep (2.5 km) V-shaped valley (Fig. 11a and b) with a low width/depth ratio (6--8, measured from HRSC dataset) that represents an outflow channel from Aram Chaos to Ares Vallis \citep{Pacifici2009}. 
The inlet of the valley along the Aram Chaos boundary is characterized by a high number of relatively small and deep channels and radial and cross-cutting grooved terrains overlying the fractured and knobby units (Figs 11b and 12).
The distal part of the inlet stands at a higher elevation with respect to part of the Aram Valley floor (Fig. 11c). This complex structure can be interpreted as an erosive remnant generated by flow converging into the channel and resulting in cross-cutting converging ridges and channels extending into the lake just upstream of the outflow point. The flow converges in the middle of the erosive remnant to create a deeply incised scour hole (Figs 11c and 12). At this location, the valley is narrower than further downstream. The different elevations of the radial grooves suggest progressive erosion of the valley inlet.
Initially, the valley floor was at a higher elevation than the chaos floor. With the progressive erosion of the inlet and valley floor, the difference in elevation decreases, whereas certain older ridges and channels are abandoned as the channel incises further until the valley reaches the present-day elevation. These observations are surprisingly consistent with analogue experiments of catastrophic, single-event crater outflows \citep{Marra2014}, including the formation of a converging erosive remnant composed by multiple cross-cutting ridges and channels.

The valley slope, obtained by removing from the profile two landslides that occur along the northern rim of the Aram Valley, is quite constant with a gentle gradient (0.004, measured from HRSC dataset) toward Ares Vallis (Fig. 11c). The present-day valley slope is lower than the initial slope prior to incision suggested by the profile along the north and south rims (0.047 for the rim N and 0.028 for rim S, measured from HRSC dataset) of the outflow valley (Fig. 11c). In the valley cross-sections 1 and 2 (Fig. 11b), abandoned flow terraces are visible on the northeastern part of the Aram channel, and their depth below the surrounding plateau (from 230 to 520 m) is a reasonable estimate of the channel water depth.

The largest portion of the Aram Chaos terrain stands at a higher elevation compared to the upper level of the Aram Valley floor (-4000 m, Figs 11 and 13) with the exception of certain basins that can reach a depth below the lowest part of the valley ($\le$ -4300 m, Fig. 13), particularly at the southern edge of Aram Chaos. The differences in elevation between the chaotic area and the Aram Valley floor, alley slope and flow convergence zone support the hypothesis of outflow and erosion processes from Aram Chaos to Ares Vallis (Fig. 11c).

\begin{figure*}[ht!]
\centerline{\includegraphics[width=0.8\textwidth]{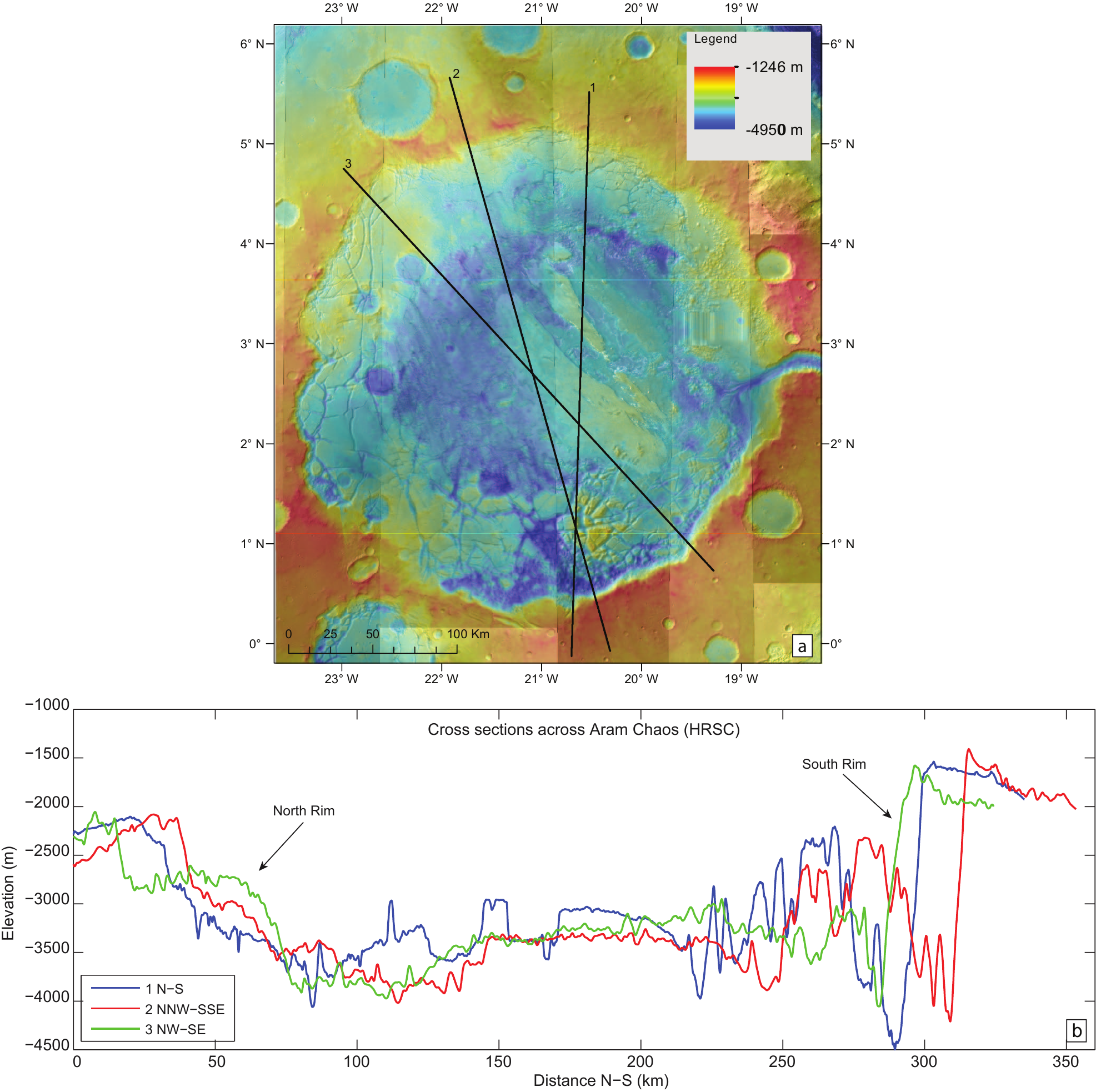}}
\caption{a)  Location of three N-S cross-sections across Aram Chaos on DEM map. b) Cross-sections across Aram Chaos: a more abrupt escapement characterizes the southern rim of the Aram crater with respect to the northern part, which presents a more gradual slope of the rim (black arrows). HRSC datasets: orbit H0401\_0001 and H0926\_0000.}
\label{collapse}
\end{figure*}

To understand why the Aram Valley is located where it is found today, we analyzed the elevation pattern of the Aram Chaos rim. At Aram valley, the crater rim before the incision of the valley was likely at approximately -1500 m (Fig. 11b, cross-section 3). The elevation map of Aram Chaos (Fig. 13) reveals the occurrence of lower elevation sections of the crater rim, especially in the northern and north-western side, which could represent a favorable outflow path way. However, no outflow remnants were found on the rim, with the exclusion of two small channels along the north-eastern side (see section 3.1 above). The occurrence of a pre-existing channel carved from the Ares Vallis during the early stages of its formation \citep[3.6 Ga,][]{Warner2009} and reused for the outflow from the Aram Chaos seem unlikely; the higher elevation of the western rim of Aram Chaos with respect to the surrounding terrain and present-day direction of the Aram Valley are incompatible with this interpretation (Fig. 13).

\begin{figure*}[ht!]
\centerline{\includegraphics[width=0.75\textwidth]{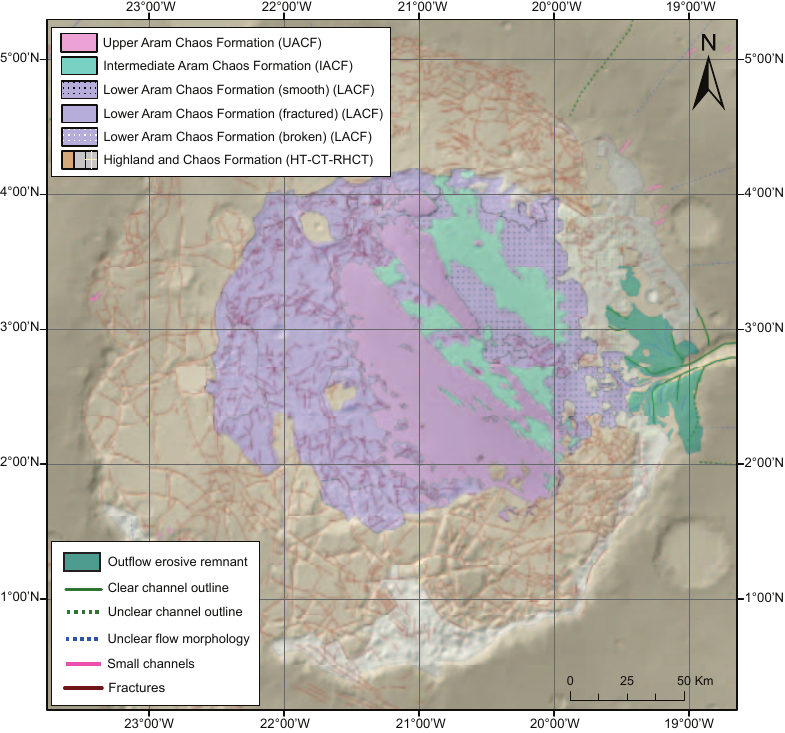}}
\caption{Structural map of Aram Chaos after \citet{Oosthoek2007a}.}
\end{figure*}

\begin{figure}[ht!]
\centerline{\includegraphics[width=\columnwidth]{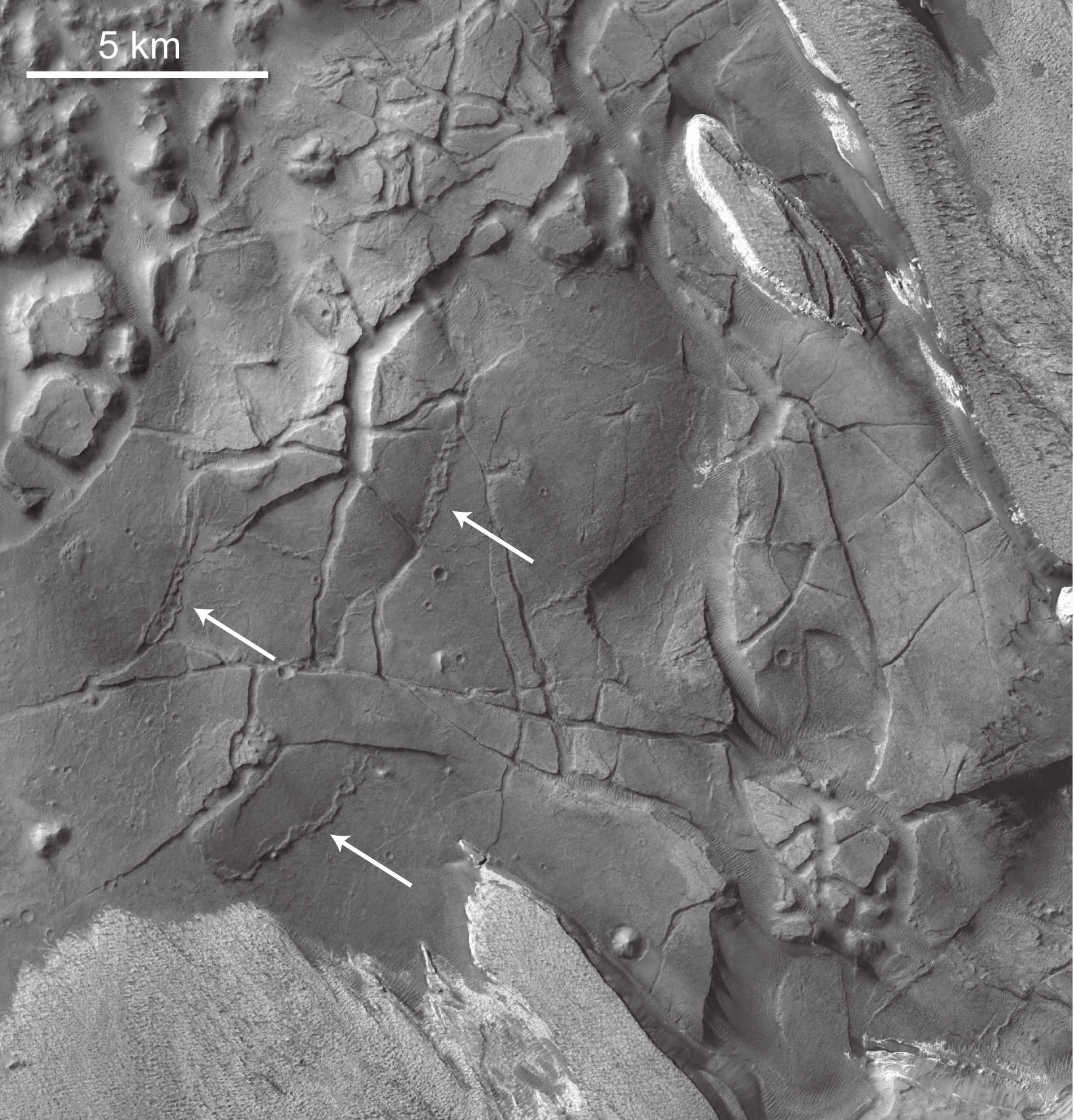}}
\caption{Small thrusts at the base of the faults rim indicated by white arrows. CTX image P19\_008311\_1836\_XI\_03N021W, 6 m/pixel.}
\end{figure}

As an alternative to surface run-off, outbursts of groundwater from a pressurized aquifer before the collapse of the sediment cap may explain the breaching of the eastern rim. Pre-existing buried fractures concentrated along the eastern rim may have been used by the water as upwelling conduits. The pressurized water would generate overpressure in the sediment cap and cause fracturing and outbursts along the eastern rim. However, the fractures generated by impact should be developed symmetrically along the entire crater rim, especially for a circular-shape crater such as Aram crater. Several outbursts would then be expected along the entire crater boundary. 

However, the eastern rim at the current location of the Ares Vallis is characterized by a shorter distance between the inner wall of the pristine crater and topographic low lying terrain (Fig. 13). The W--E profile along Aram Chaos (Fig. 14) indicates that the Western Valley \citep{Pacifici2009}, that surrounds the western part of Aram Chaos and the tributary of Hydapsis Chaos has an elevation comparable to that shown by Ares Vallis before the Aram Valley outflow event, such as -3500 m. This elevation, however, was likely obtained by the last flood sourced from Hydapsis Chaos and dated at 2.6 Ga \citep{Warner2010a}. The first outflow event carving the Hydapsis tributary was coeval with the first event of Ares Vallis \citep{Warner2010a}, and the erosive surface was most likely located at -2600 m (Fig. 14a). 
The steepest subsurface gradient was likely localized between Aram Chaos and Ares Vallis (0.018--0.023 compared to 0.012--0.016 of the western part), which most likely made the groundwater flow towards Ares Vallis with the outlet located between -3000 and -3500 m (Fig. 14). It is therefore likely that the outflow started as groundwater seepage and evolved into an outflow valley by groundwater-sapping and headcutting from Ares Vallis through the crater rim \citep{Howard1988,Brocard2011}. After the breaching of the crater rim, the process would have developed much faster to evolve as a massive outflow, and the channel would have deepened rapidly to complete the carving of Aram Valley. The observations and analogue modeling support the hypothesis of a single and continuous outflow carving the valley, although subsequent small outflow events may have occurred because of groundwater upwelling.

As observed by \citet{Warner2009}, the Aram Valley is graded to the final erosive surface of Ares Vallis with no evidence of intersecting or truncating flood grooves or a knickpoint occurrence at the confluence. Furthermore, the topography of Ares Vallis (Fig. 15) shows an abrupt increase in slope at the confluence with Aram Valley as well as an increase of width. These observations clearly support the interpretation of \citet{Warner2009}, who suggested that the water outflow from the Aram Valley was synchronous with the final erosive event of Ares Vallis (approximately 2.5 Ga).

It is now possible to infer a relative chronology of the Aram Chaos outflows, which started with two relatively small channels along the north-eastern part of the rim and continued in the larger Aram Valley as massive outflow once the groundwater headcutting was completed. This process was active for a relatively short period and was coeval with the late stages of Ares Vallis formation.

\section{Outflow time scale and water volume estimates}
In this section, we describe the method for deriving flow volume and formative time scales from the morphology and morphometric characteristics of outflow channels. A reconstructed flow rate will be combined with the reconstructed volume of water to derive an estimate of event duration. The rate of sediment removal calculated on the basis of the reconstructed flow rate will be combined with the volume of the valley to derive a second, independent estimate of event duration, which we will then compare to the first estimate to arrive at a best estimate of the event duration. Sources of uncertainty are discussed in combination with a likely course of events in the excavation of the channel.

\subsection{Time scale and water volume determination}
The principle of the calculation of formative time scale is that a flow requires a certain time to remove or deposit a known volume of sediment \citep{Kleinhans2005}. 
The volume of sediment eroded from the valleys is estimated from cross-sections and the length of the valleys. The estimates of cross-section surface area, averaged over several of HRSC profiles, have been multiplied by the length of the valleys (Figs 10 and 11). This yields an average volume of 460 km$^3$ for the Aram Valley and 20 and 12 km$^3$ for channels 1 and 2, respectively (Table 1).

The flow flux from the Aram Chaos crater was likely (nearly) clear water that resulted from ponding after the collapse so that the sediment transport capacity of the flow was entirely available for erosion of the channels. This clear water scour was basically the inverse of the deposition of crater lake deltas from a sediment-laden flow that enters a crater lake \citep{Kleinhans2005,Kraal2008}. The sudden transition from sediment transport to zero transport in the delta case and vice versa in the channels case allows us to calculate the time scale [$T_s$] of formation directly from the volume of displaced sediment [$V$ = 460 km$^3$] and the sediment transport rate [$Q_s$] (corrected for porosity) as $T_s=V/Q_s$ \citep{Kleinhans2005}. The sediment transport rate is calculated from the flow flux through the channel. 

Flow flux is calculated by the following steps. First, the width, flow depth and gradient of the channel are estimated, and the width, cross-sectional valley shape and valley gradients are estimated from HRSC topography (Figs 10 and 11).
Maximum flow depth is estimated from terraces heights, and hydraulic roughness is then calculated,
\begin{equation}
f=\frac{8}{(2.2(\frac{h}{D_{50}})^{-0.055}S^{-0.275})^{2}}
\end{equation}
from which the flow velocity 
\begin{equation}
u=\sqrt{\frac{8ghS}{f}}
\end{equation}
and flow discharge 
\begin{equation}
Q_w=uhW
\end{equation}
follow. $h$ is the channel depth, $D_{50}$ is the median grain size, S is the channel slope, $g$ is the martian gravity and $W$ the channel width. The water depth inferred from terraces ($h$) is within the expected range based on the resulting width-depth ratio of the flow (20 for narrow terrestrial gravel bed rivers) and results in reasonable Froude numbers and sediment mobilities (expressed as non-dimensional Shields number, Table 1) \citep{Kleinhans2005}.

To estimate the total water volume $V_w$ that must have come out of the Aram Chaos crater to form the observed channels, the formative time scale $T_s$ for channel excavation can be multiplied by the flow flux $Q_w$ so that $V_w=T_s Q_w$ \citep{Kleinhans2005}. This yields a water volume estimate of 9.3$\cdot$10$^4$ km$^3$ for the Aram Valley and 2.3$\cdot$10$^3$  and 2.1$\cdot$10$^3$ km$^3$ for the two small channels.

Sediment flux is calculated by two methods: one assuming a bed load-dominated transport (with mostly rolling and saltating particles and limited energy) and one assuming a suspended load-dominated event (non-cohesive granular material). The sediment mobility, which depends on flow and sediment properties, is used to determine which of the two transport modes is valid. A recent granular debris flow in the middle of the present channel (Fig. 16) indicates that failure of the side walls would immediately collapse the material into non-cohesive granular sediment. This debris flow formed as a dry granular flow because only this rheology allows for a debris flow that goes slope-upward on the opposite valley wall. This is evident in that the material is neither cohesive nor strongly lithified in addition to the extensive analysis in \citet{Grotzinger2012},which allows the application of transport capacity predictors. Furthermore, these predictors are reasonably accurate in cases of weakly lithified material \citep[see discussion in][]{Kleinhans2005}.

Classical sediment transport capacity predictors are employed corrected for gravity, and sediment properties such as estimated in \citet{Kleinhans2005} are used. The ratio of suspended and bed load transport is much larger than unity, so that the system is suspension dominated and the appropriate predictor is used \citep{Kleinhans2005}.
The volumetric transport rate [m$^{3}$/s] is
\begin{equation}
Q_s=\frac{1}{1-n }\Phi _s\sqrt{Rg}D_{50}^{3/2}W
\end{equation}
where
\begin{equation}
\Phi _s=\frac{0.4}{f}\theta ^{2.5}
\end{equation}
is the non-dimensional suspended load transport predictor, $\theta$ is the shear stress, $n$ is the porosity, $R$ is the relative submerged density.
The maximum time scale for channel formation ($T_s=V/Q_s$) then becomes of the order of tens of days (Table 1 - Formative time scale) with an order of magnitude in the uncertainty range \citep{Kleinhans2005}.

\begin{figure*}[ht!]
\centerline{\includegraphics[width=0.75\textwidth]{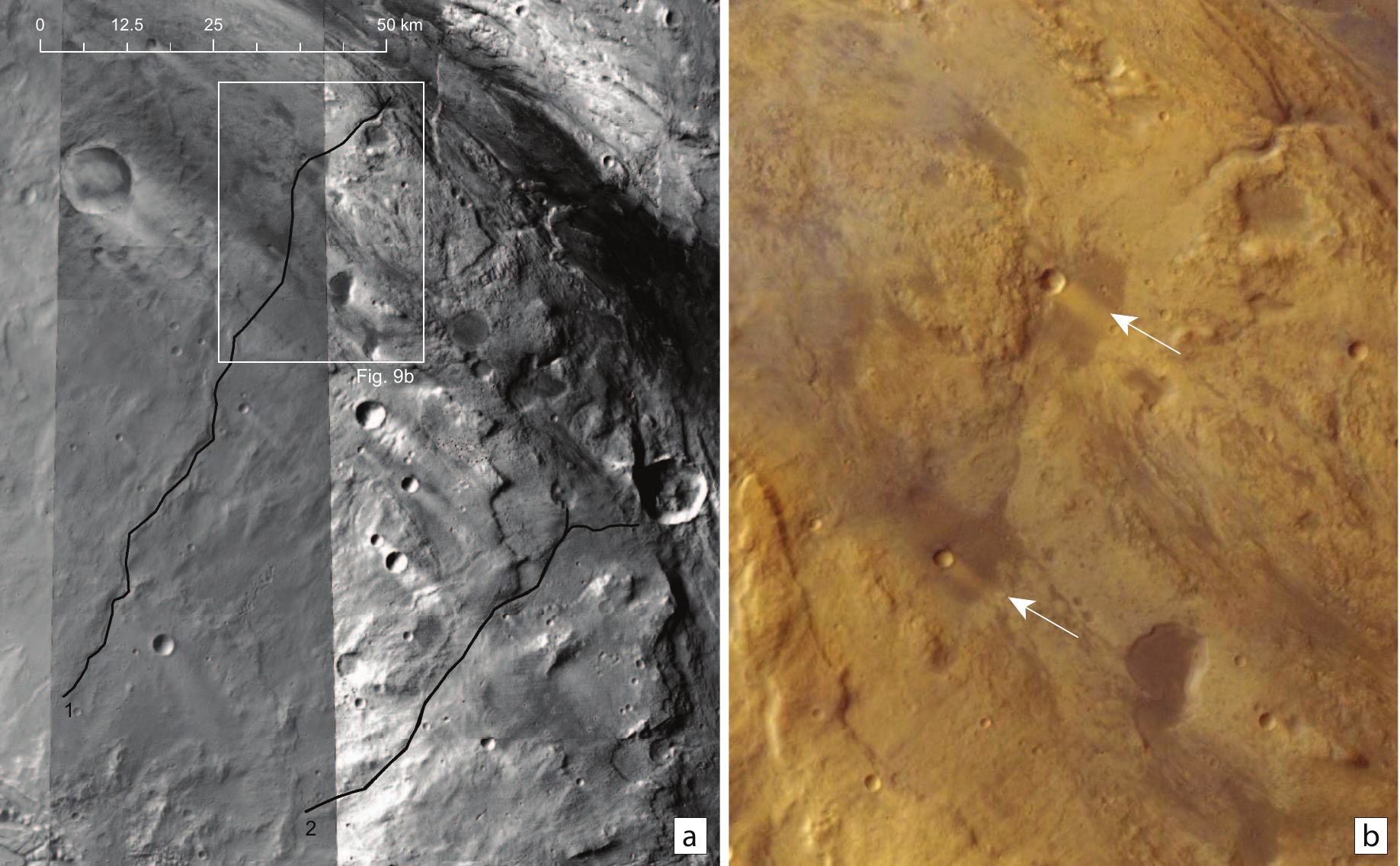}}
\caption{Visible images of north-eastern rim of Aram Chaos. a) Two small outflow channels are traceable across the rim until Ares Vallis. They are likely cut by the late stages of Ares Vallis outflow. HRSC dataset, orbits H0934\_0000, H0945\_0000 and H0401\_0001 b) Detail of the lower part of northern channels: the channel floor is truncated by the ejecta blankets of two recent impact crates but it is still traceable until its termination in the lower part of Ares Vallis (upper right corner). Visible image, credit Google Mars.}
\end{figure*}

\begin{figure}[ht!]
\centerline{\includegraphics[width=.9\columnwidth]{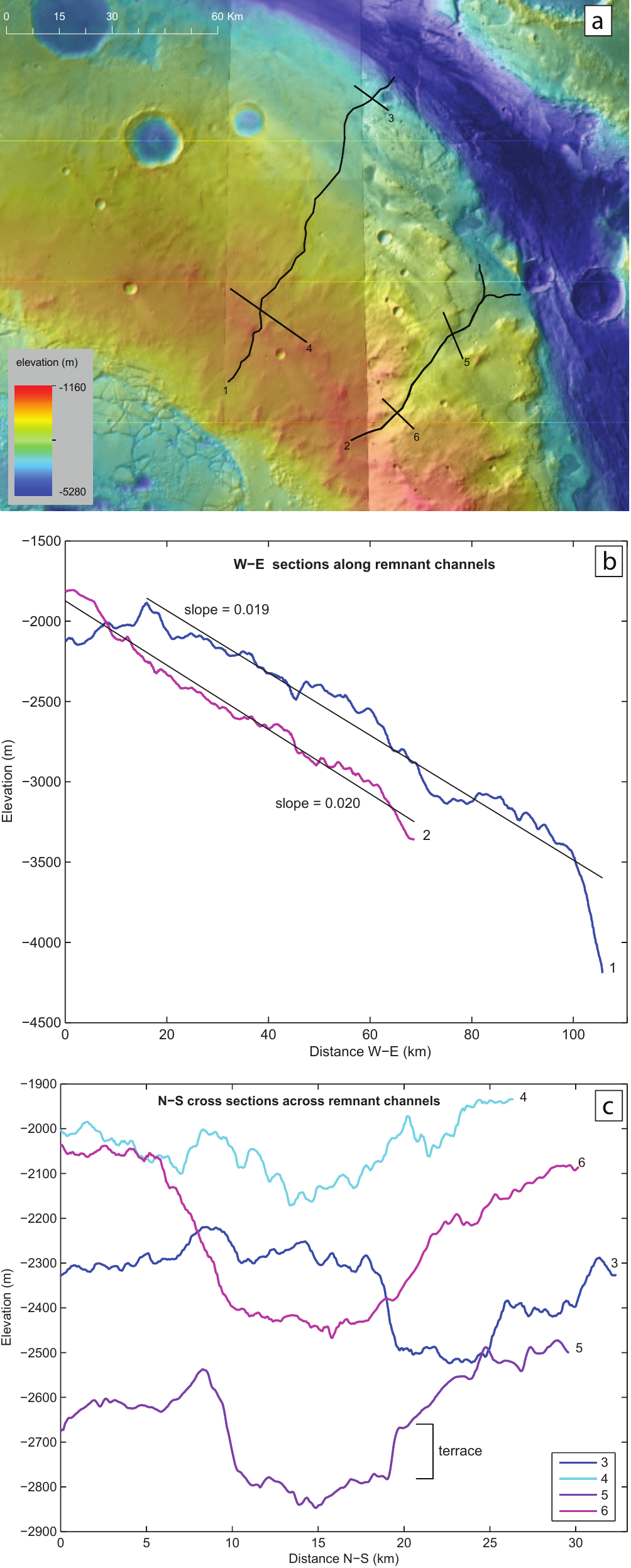}}
\caption{a) Location of the cross section of small outflow channels on DEM map. b) W--E profiles along the channels. They show a rather constant and similar slope and their outlets are truncated by the late erosional features of Ares Vallis \citep{Warner2009}. c) N--S cross-sections across the outflow channels. The abandoned terrace identified in the section 5 referred to the southern channels (2) is used as an estimate of channel depth.
HRSC dataset, orbits H934\_0000 and H945\_0000.}
\end{figure}

\begin{figure}[ht!]
\centerline{\includegraphics[width=1\columnwidth]{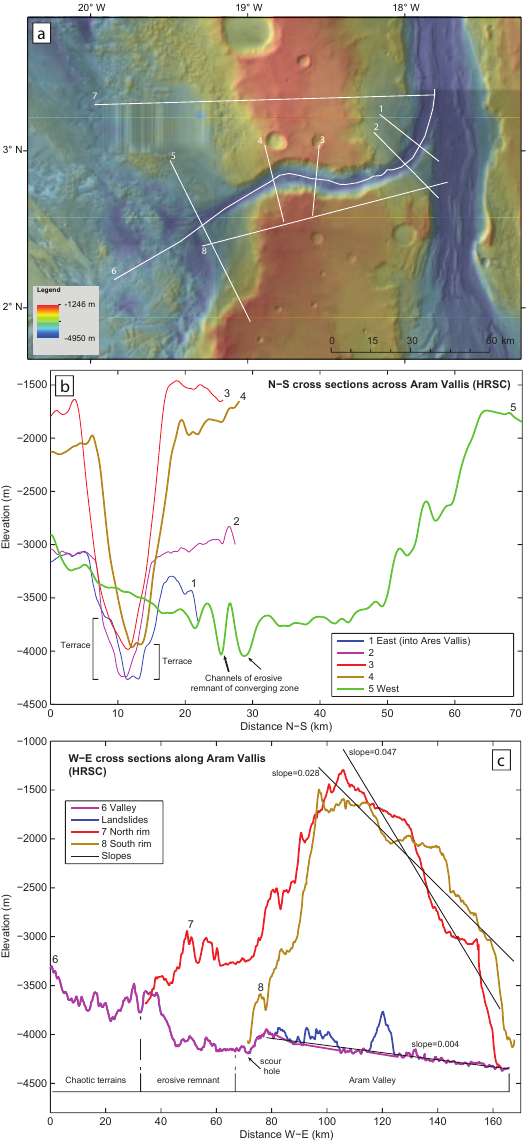}}
\caption{a) Location of the cross-sections of Aram Valley on DEM map. b) N--S cross sections across Aram Valley. In the sections 1 and 2, obtained near the valley outlet, some abandoned flow terraces are visible and their depth below the surrounding plateau (from 230 to 520 m) is interpreted as an estimate of channel depth. c) W--E cross-sections along the Aram Valley and surrounding rims. The present-day valley slope (section 6) is lower than the pristine slope suggested by the profile along the north and south rims of the outflow valley (sections 7 and 8). HRSC datasets from orbit H0401\_0001 and H0923\_0000.}
\label{hrsc}
\end{figure}

\subsection{Analysis of uncertainties}
For their morphometry and position, the two small channels are treated as representative scenarios for the early outflow stages. For the Aram Valley, instead, various scenarios for the event are presented in Table 1 and cover the likely range of conditions, formative time scales and volumes involved.
Five different scenarios are presented and referred to as follows: (i) a best guess scenario, where the best guesses were used for channel geometry and gradient and for the outflow volume from the crater and eroded sediment volume from the channel; (ii) a slower scenario, where a larger estimate for water depth and smaller estimate of gradient were combined to produce the smallest likely flow velocity and bracket the likely time scales; (iii) a faster (larger gradient) and smaller (lower water depth) scenario representative of the initial stages of aerial water flow; (iv--vii) the best guess scenario combined with the larger and smaller water volume estimate or the larger and smaller sediment volume, also bracketing the likely time scales. The comparison indicates a maximum formative time scale lower than 30 days and a water volume greater than 1$\cdot$10$^4$ km$^3$.

\begin{figure}[ht!]
\centerline{\includegraphics[width=\columnwidth]{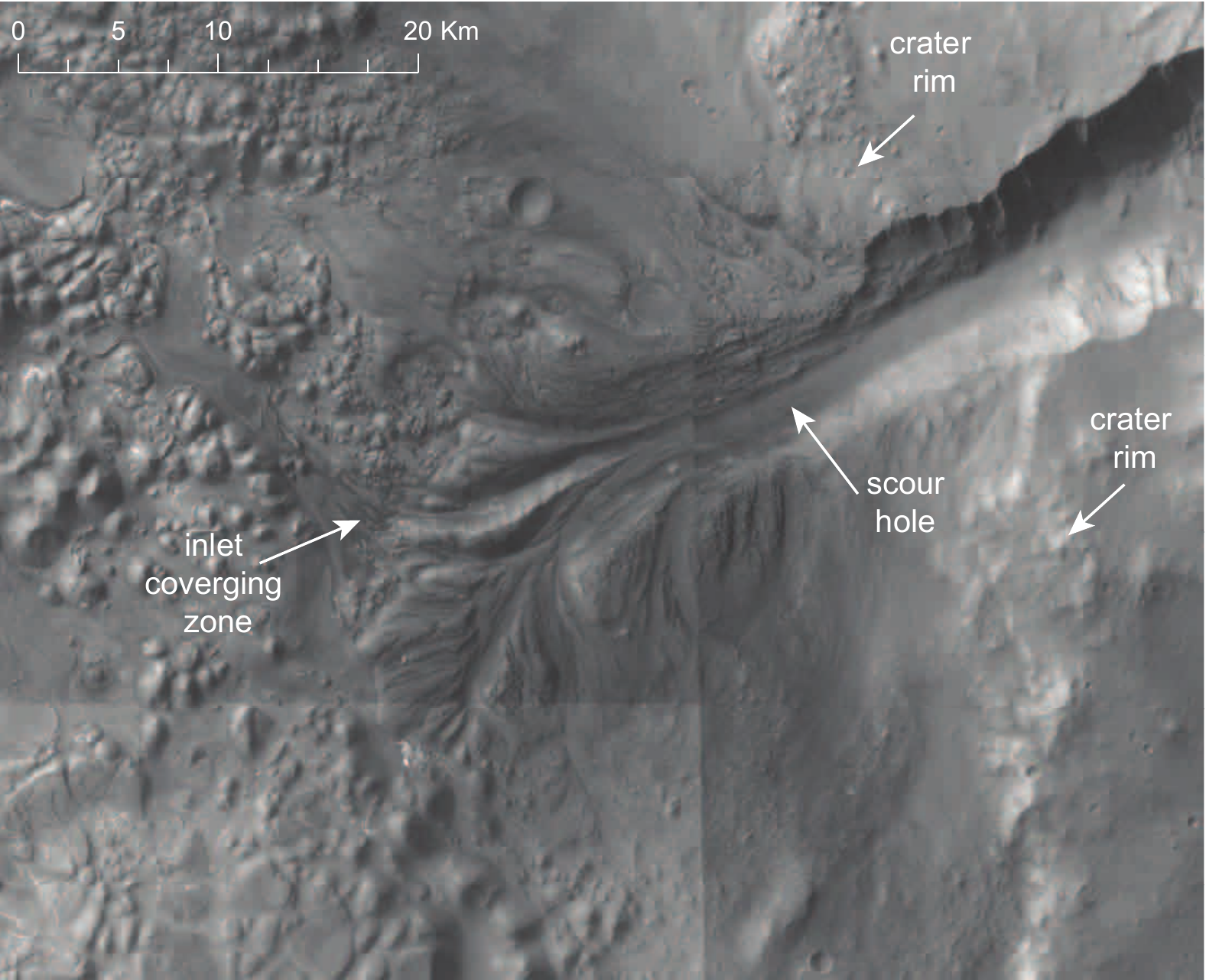}}
\caption{Erosive remnant of the flow convergence zone located at the inlet of Aram Valley. It is characterized by a high number of relative small and deep channels and radial grooved terrains overlying the fractured and knobby unit. This structure suggests a converging flow into the valley. HRSC dataset, orbit H945\_0000.}
\label{sections}
\end{figure}

Furthermore, a sensitivity analysis was performed for the Aram Valley for the parameters affected by significant uncertainties and was performed for water depth in particular, but also for valley shape and valley gradient (Fig. 17). The resulting time scales for the water flow and sediment removal are plotted against nine water depths for 3 valley gradients and 3 sediment volumes that reflect three different valley shapes. Because it is the water flow that removes the sediment, the two time scales must be  equal ($T_w=T_s$) and the measured amount of water must be exactly sufficient to erode the measured amount of sediment based on predicted flow and sediment transport rates.

The result is remarkably consistent (Fig. 17): the formative time scales for release of water and removal of sediment are equal at tens of days for a water depth range of approximately 250--400 m, which takes into account all the conservative estimates of uncertainties. This water depth is in accurate agreement with observed terrace heights in the valley (Fig. 11b), which are indicative of a water depth that was found in the experiments of \citet{Marra2014}. The order of magnitude of the uncertainty in time scale for sediment removal $T_s$ results from the uncertainty of water surface gradient (0.004), which was estimated from the final bed surface gradient (ignoring later debris flow deposits) and maximum likely slope from the crater rim surface just outside the channel (0.02) after the breaching. Neither is likely to be representative for the entire period, so we selected a gradient magnitude in between for the best guess scenario (0.0045). The estimated uncertainty of an order of magnitude also includes the contributions of the chosen friction relation and the total load sediment transport predictor.

\begin{figure}[ht]
\centerline{\includegraphics[width=\columnwidth]{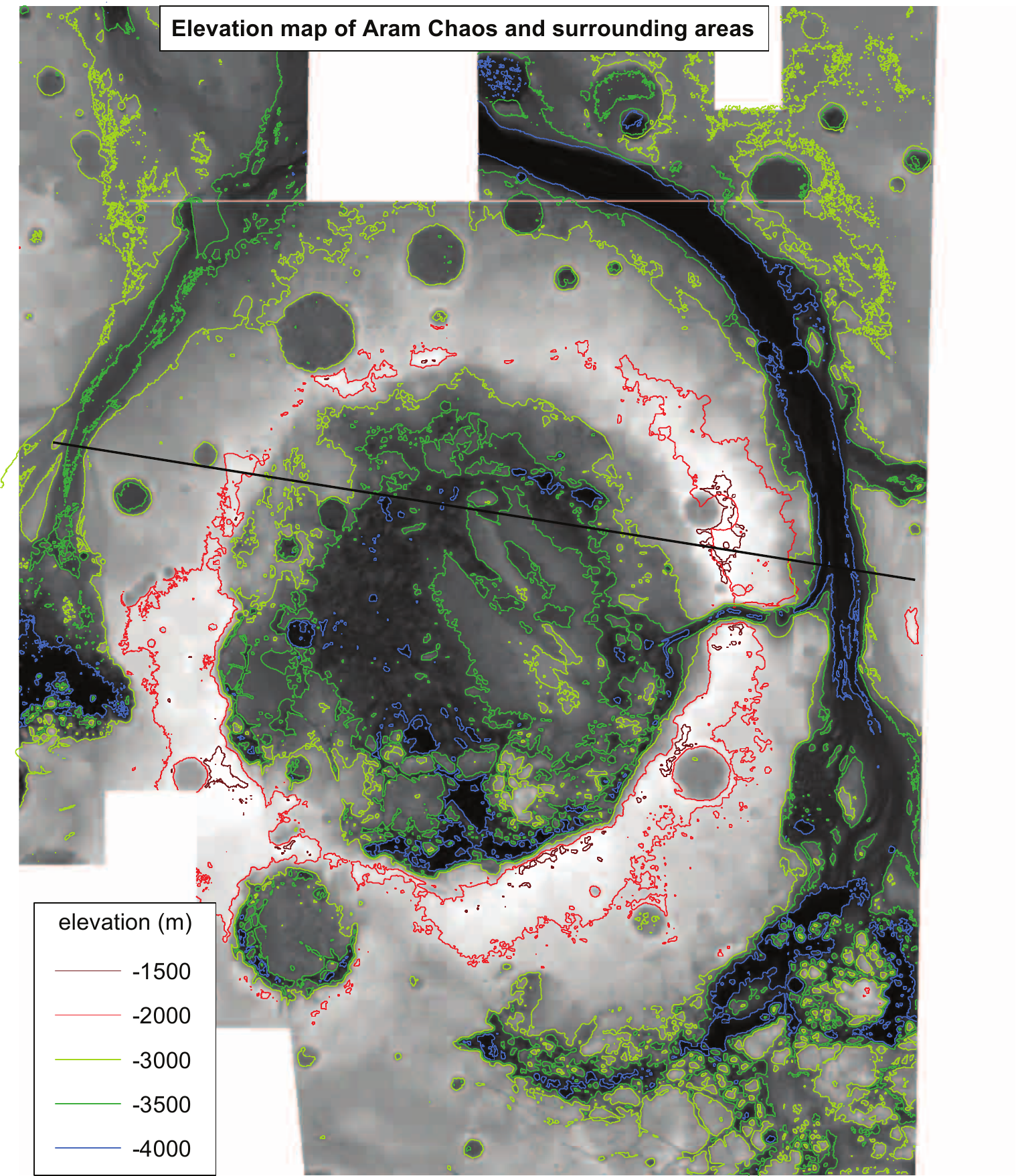}}
\caption{Elevation map of  Aram Chaos - Valley system: five topographic surfaces are highlighted, from -1500 to -4000m. The most part of the Aram chaos terrain stands at higher elevation compare to the upper level of Aram Valley floor (-4000 m). The highest part of the Aram crater rim is located around the Aram Valley. This area is also characterized by a steep gradient between the inner and outer wall of the pristine crater (distance between inner crater rim and greenish surrounding surfaces). HRSC datasets from orbits H0401\_0001 and H0923\_0000.}
\end{figure}

\section{Comparison with evolutionary scenarios}
From the analysis performed on Aram Chaos and Aram Valley it is clear that four main elements are important for the validation of the proposed evolutionary scenarios of Aram Chaos and chaotic terrains in general: fracturing, the amount of subsidence of the chaotic terrains, total water volume and water flux (i.e., time scale) of the outflow channels (Table 2). For Aram Chaos, at least 1.5 km of collapse is estimated by the structural and morphological analysis. The Aram Valley outflow channel was likely carved with a large amount of water in less than 30 days and in a maximum 1 or 2 outflow events.

The first group of proposed mechanisms (Fig. 1) accounts for the interaction of volcanic activity with the cryosphere \citep[e.g.,][]{Ogawa2003,Meresse2008}. Although these models could explain the occurrence of hydrothermal minerals in terms of magma/ice interactions \citep{Ogawa2003}, the general lack of evidence of magmatic activity within the chaotic terrains raises questions about the applicability of the model. However, the relative proximity of the Tharsis region with the Xanthe Terra chaotic region could provide evidence of the occurrence of active volcanism in that chaotic terrain.
 In those models, the subsidence is partially related to the volume loss by melting of the ice in the pore space of the cryosphere and the water release. This implies that by using an extremely high constant porosity of 20$\%$, at least 7.5 km of cryosphere should melt and discharge to achieve a subsidence of 1500 m, as observed for the Aram Chaos. Using a much more reliable depth--porosity relation \citep{Clifford2010} and taking into account a very high surface porosity of 30$\%$, the maximum amount of subsidence achievable for 20 km of thick cryosphere is less than 1 km.
Although the water volume and water flux generated by the volcanic-cryosphere model is not estimated, the amount of water released from the cryosphere, which depends primarily on porosity and permeability, is low \citep[10$^0$--10$^2$  km$^3$, 10$^3$  km$^3$ only for extreme values of permeability,][]{Harrison2008} in comparison with the volume of water required to carve the outflow channels (9.3$\cdot$10$^4$ km$^3$ for Aram Valley). Only with a large number of outflow events is it possible to archive a comparably large water volume. This implies a continuous recharging and freezing of cryosphere and repeated intrusions. Furthermore, considering the low mean discharge suggested by the numerical modeling of the martian aquifer \citep[10$^3$--10$^4$  m$^3$/s,][]{Hanna2007a}, the duration of the outflow events \citep[10$^0$--10$^2$ years,][]{Hanna2007a} is not compatible with the catastrophic event (tens of days) suggested by the hydrologic analysis performed for the Aram Valley.

\begin{figure*}[ht!]
\centerline{\includegraphics[width=.9\textwidth]{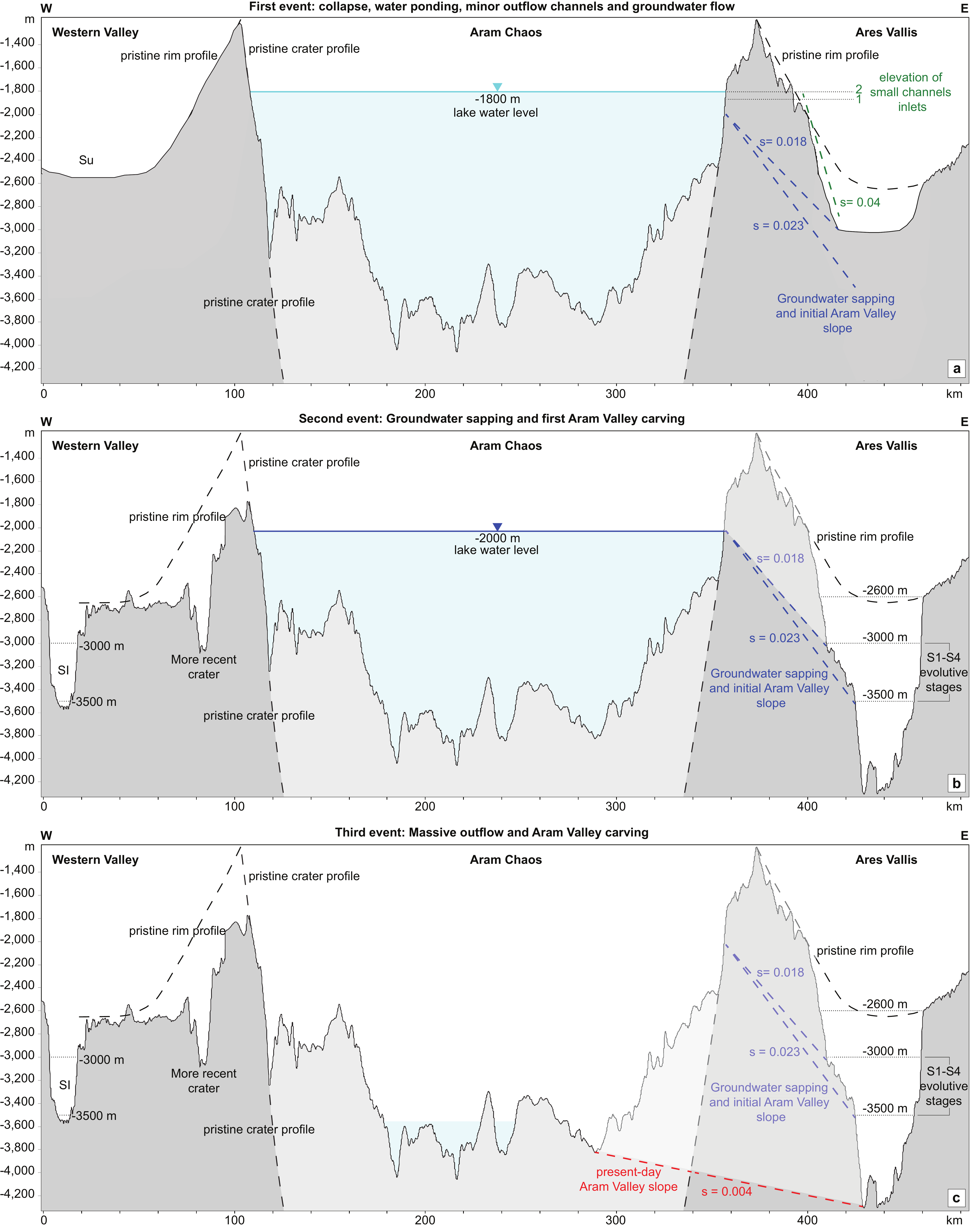}}
\caption{W--E profile along the Aram Chaos (see Fig. 13 for the location) and interpretation of the events involved in the buried sub-ice lake scenario. Dashed lines represent the slopes of outflow channels (green and red) and groundwater (blue). Black dotted lines indicate the base levels of Ares Vallis before (-2600 m) and after (-3000 and -3500 m) the first erosive events \citep[S1--S4,][]{Warner2009}. SU and SI are erosive events of Western Valley \citep{Warner2010a}. With dashed black lines the pristine crater and rim profiles are indicated. With transparency the progressive erosion of Aram Valley is shown.
a) After the collapse water ponding, first outflows and groundwater flow along the eastern rim occur; b) The water level drops to -2000 m, the small outflows end and water sapping starts the erosion of Aram Valley; c) Once the breaching of the eastern rim occurs, massive water outflow completes the carving of Aram Valley up to the present-day morphology.}
\end{figure*}

For the second mechanism (Fig. 1), the aquifer model \citep[e.g.,][]{Carr1979,Clifford1993,Hanna2007a,Harrison2009}and numerical and analogue modeling suggests that a very high number of outflow events are required to achieve the amount of water sufficient to carve the outflow channels. The catastrophic characteristics of the outflow mechanism and its formative time scale suggested by the Aram Valley analysis are in conflict with this interpretation.
The water flood from a subsurface aquifer initiates when superlithostatic pore pressures within a confined aquifer lead to the propagation of hydrofractures through the confining cryosphere to the surface \citep{Hanna2007a}. This could explain the fractures occurring in the chaotic terrains. However, the mechanism of hydrostatic fracturing has not been directly implemented in the hydrologic models.
Furthermore, the collapse and subsidence over at least 1500 m that was simultaneous with fracturing affecting the chaotic terrains region is not compatible with hydrofracturing because only minor subsidence and collapse is acceptable if the flow properties of the aquifer are to be maintained to achieve multiple outflow events. In this model, the high subsidence observed in the chaotic terrains area is treated as a pre-existing topographic low along which the hydraulic head can reach the lithostatic pressure and lead to fracturing. Morphological analysis of Aram Chaos, instead, suggests that the subsidence, which is mainly taken up by the rim fault (1500 m displacement), is coeval with overall fracturing of the chaotic terrain.

\begin{figure*}[ht!]
\centerline{\includegraphics[width=0.7\textwidth]{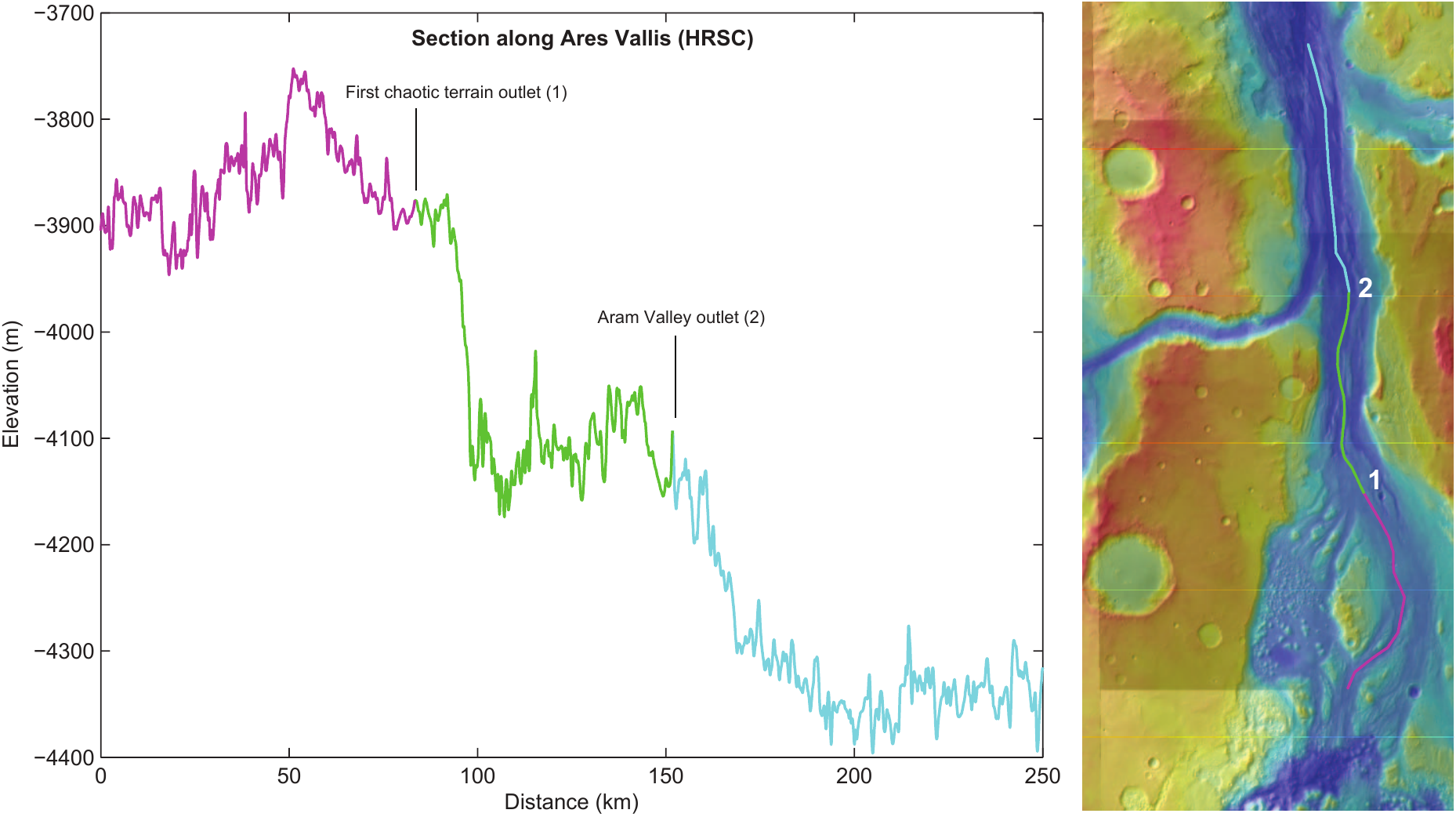}}
\caption{N--S profile along Ares Vallis: when the main valley floor crosses the Aram Valley the slope increases as well as the valley width (transition from green to cyan lines). This suggests that the water outflow from the Aram Valley was synchronous with final erosive event of Ares Vallis. HRSC dataset, orbits H923\_0000 and H934\_0000.}
\end{figure*}

\begin{table*}
\centering
\caption{Calculation of formative time scale for channel excavation and the necessary water volume involved (to be compared to the geologically estimated water volume) for different scenarios. Slower = deeper and small gradient; Faster = shallower and large gradient. See the italic values for the variables changed within each scenario. The friction factor is obtained using the equation 13 of \citet{Kleinhans2005}.}
\footnotesize
\begin{tabular}{llcccccccc}
\hline

Calculation&Parameters&Units& \multicolumn{7}{c}{Scenario}\\

&&&Best guess&Slower&Faster &Less water&More water&Less sed.&More sed.\\

\hline
Input&Channel width&m&8000&8000&8000&8000&8000&8000&8000\\

&Channel depth&m&375&\it{520}&\it{230}&375&375&375&375\\

&Slope&m/m&0.0045&\it{0.004}&\it{0.02}&0.0045&0.0045&0.0045&0.0045\\

Discharge calculation&friction factor&-&0.21&0.20&0.45&0.21&0.21&0.21&0.21\\

&Velocity&m/s&14.8&16.5&17.0&14.8&14.8&14.8&14.8\\

&Froude number&-&0.40&0.37&0.58&0.40&0.40&0.40&0.40\\

&Discharge&m3/s&4.5$\cdot$10$^7$&6.8$\cdot$10$^7$&3.1$\cdot$10$^7$&4.5$\cdot$10$^7$&4.5$\cdot$10$^7$&4.5$\cdot$10$^7$&4.5$\cdot$10$^7$\\

Bedload transport&Shields parameter &-&0.48&0.56&0.71&0.48&0.48&0.48&0.48\\

&Nondim. transport rate&-&2.45&3.07&4.44&2.45&2.45&2.45&2.45\\

&Volum. transport rate&km3/day&0.20&0.26&0.37&0.20&0.20&0.20&0.20\\

&water/sediment ratio&-&1.9$\cdot$10$^3$&2.3$\cdot$10$^3$&7.3$\cdot$10$^3$&1.9$\cdot$10$^3$&1.9$\cdot$10$^3$&1.9$\cdot$10$^3$&1.9$\cdot$10$^3$\\

Total load transport&Shields parameter&-&6.7&8.3&18.4&6.7&6.7&6.7&6.7\\

suspension dominated&Nondim. transport rate&-&2.3$\cdot$10$^2$&3.9$\cdot$10$^2$&1.3$\cdot$10$^3$&2.3$\cdot$10$^2$&2.3$\cdot$10$^2$&2.3$\cdot$10$^2$&2.3$\cdot$10$^2$\\

&Volum. transport rate&km3/day&19&33&108&19&19&19&19\\

&water/sediment ratio&-&2.0$\cdot$10$^2$&1.8$\cdot$10$^2$&2.5$\cdot$10$^1$&2.0$\cdot$10$^2$&2.0$\cdot$10$^2$&2.0$\cdot$10$^2$&2.0$\cdot$10$^2$\\

Erosion&sediment valley volume&km3&4.5$\cdot$10$^2$&4.5$\cdot$10$^2$&4.5$\cdot$10$^2$&4.5$\cdot$10$^2$&4.5$\cdot$10$^2$&\it{3.7$\cdot$10$^2$}&\it{5.4$\cdot$10$^2$}\\

Measured flood&outflow volume&km3&9.2$\cdot$10$^4$&9.2$\cdot$10$^4$&9.2$\cdot$10$^4$&\it{6.2$\cdot$10$^4$}&\it{1.2$\cdot$10$^5$}&9.2$\cdot$10$^4$&9.2$\cdot$10$^4$\\

\bf{Formative time scale} &\bf{time scale}&days&\bf{24}&\bf{14}&\bf{4}&\bf{24}&\bf{24}&\bf{20}&\bf{29}\\

\bf{Predicted volume of flood}&\bf{flood volume}&km3&\bf{9.3$\cdot$10$^4$}&\bf{8.2$\cdot$10$^4$}&\bf{1.1$\cdot$10$^4$}&\bf{9.3$\cdot$10$^4$}&\bf{9.3$\cdot$10$^4$}&\bf{7.5$\cdot$10$^4$}&\bf{1.1$\cdot$10$^5$}\\

\hline
\label{}
\end{tabular}
\end{table*}

With respect to the aquifer model, the gas (or salt) hydrated mechanism \citep{Max2001,Montgomery2005} implies the increase in pore-pressure (and the consequent cryosphere hydrofracturing) because of the gas and/or water released by salt hydrate deposits or clathrates.
However, for the same reasons as the required amount of water, time scale of floods and amount of subsidence discussed for the aquifer model, the gas (or salt) hydrated model can not explain the sequence of events in Aram Chaos. Furthermore, the large number of floods and relative time scale required by the permeability of the aquifer contrasts with the rapid degassing (or dewatering) proposed for this mechanism. Although this model could explain the occurrence of mono and poly-hydrated minerals, a model to explain gas/salt hydrated occurrence on Mars is not yet available.

In the buried sub-ice lake scenario proposed by \citet{Zegers2010}, melting of a buried ice layer followed by mechanical destabilization of the sediment cap and massive water outflow is able to explain the strong subsidence and fracturing of the chaotic terrains as well as the large amount of water required to carve the outflow channels in a catastrophic way. Analogue modeling of the process \citep{Manker1982} appears to confirm the resulting chaotic morphology, and analogue modeling of the outflow process \citep{Marra2014} confirms all diagnostic features in the resulting valley morphology. Furthermore, numerical modeling suggests that a heat flow of 25 mW/m$^2$ together with the thermal insulation of overburden is sufficient to induce relative slow melting of the ice, without the need of a magmatic event \citep{Zegers2010,Schumacher2011}.
\citet{Zegers2010} estimated the volume of liquid water that was produced in a single chaotization event by taking the subsidence as a measure of the water escaped. Given a crater diameter of 280 km and a subsidence of 1--2 km and assuming a simple cylindrical shape of the crater, a volume of 0.6--1.2$\cdot$10$^5$ km$^3$ results. The uncertainty in the calculation of the water volume results from an uncertainty in the magnitude of subsidence (between 1 and 2 km, with local variations). As a best estimate of the water volume released assuming a single event, we therefore use 9.2$\cdot$10$^4$ km$^3$ (Table 1 - measured flood volume).
The total predicted volume of flood obtained for the Aram Valley and two small channels by the hydrological analysis (9.3$\cdot$10$^4$ km$^3$, Table 1 - predicted volume of flood) is in the same order of magnitude of the independent estimate of the volume from crater geology that assume a single event.

In conclusion we propose that a buried sub-ice lake scenario can best explain the observations and the water volume and outflow time scale estimates made for the Aram Chaos--Valley system.

\section{Scenario for the evolution of the Aram Chaos--Valley system}
In this section, we use the morphological and chronological information together with quantitative estimates of the outflows (time scale and water volume) presented in the previous sections to obtain a description of the events involved during the Aram Chaos and Aram Valley outflow channel formation.
The morphological and hydrological analysis of the Aram Chaos - Valley system suggests fracturing, subsidence and collapse of terrain within the crater and a catastrophic water outflow along the eastern rim. Applying the scenario presented by \citet{Zegers2010}, Aram Chaos would be the result of destabilization of a buried sub-ice lake with a consequent collapse of the sediment cap and massive expulsion of water toward the surface, followed by temporary ponding. Two small channels were carved by surface run-off, and at the same time, groundwater flow started from the lake to Ares Vallis, generating seepage erosion. After the breaching of the Aram Chaos rim, a massive outflow of water carved Aram Valley in a very short time scale.  The formative time scale and the volume of water involved in carving the channel were calculated as if the fluxes were constant during the event. It is likely that the fluxes were not constant but rather a result of the likely process of channel formation, which are described below and can be approached with constant fluxes for our purpose.

In the following sections, we discuss in detail the evolutionary events of the  Aram Chaos--Valley system.  

\begin{figure}[ht!]
\centerline{\includegraphics[width=\columnwidth]{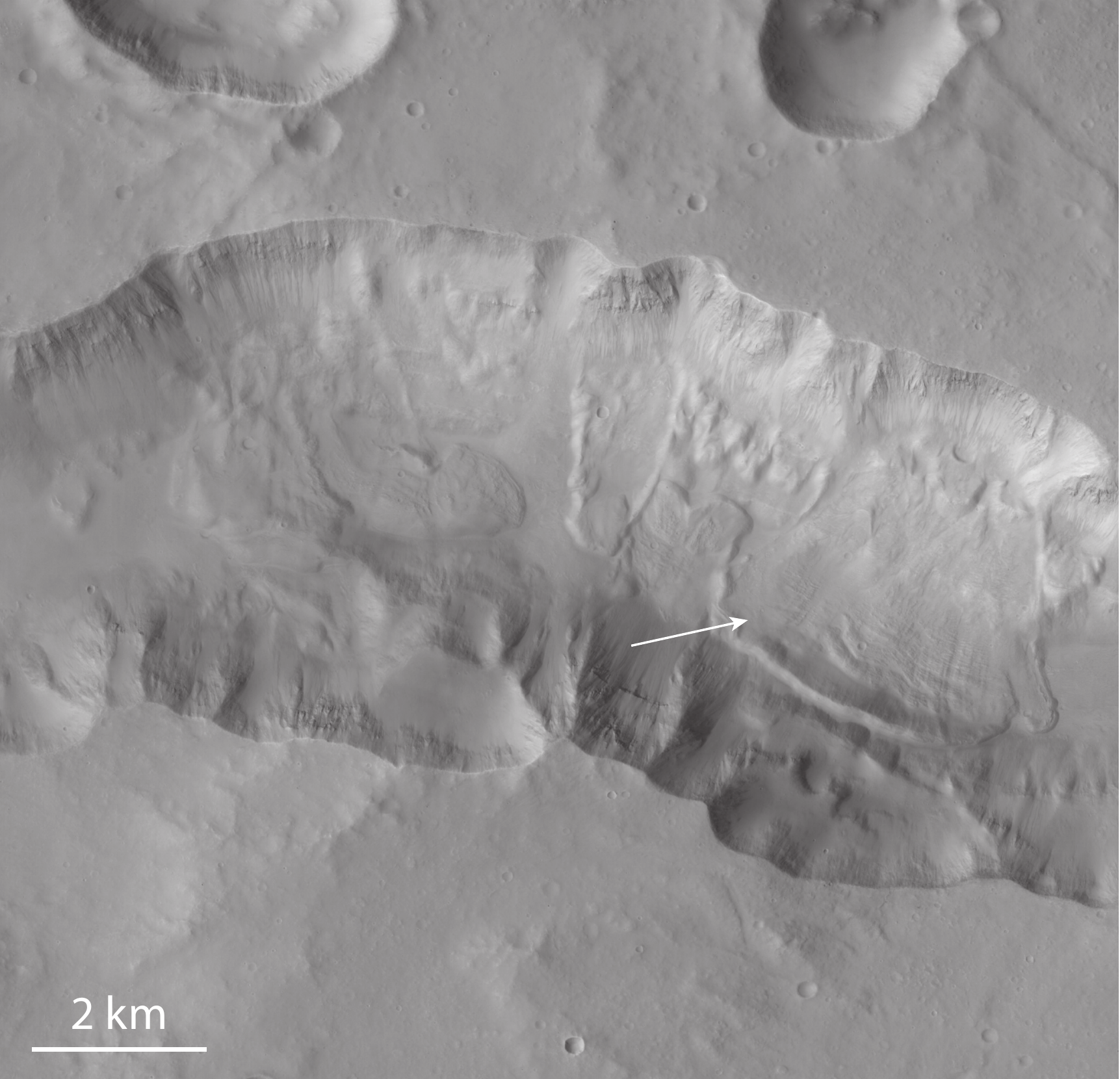}}
\caption{Details of the eastern debris flow located in the Aram Valley (white arrow). This debris flow goes slope-upward on the opposite valley wall (south wall) indicating a dry granular rheology flow. CTX image P05\_003129\_1842\_XI\_04N018W.}
\end{figure}

\subsection{Aram Chaos collapse}
Geological, morphological and structural analyses of Aram Chaos suggest a large scale collapse and subsidence of the entire area of 1000--1500 m, which is compatible with a massive expulsion of liquid water from the subsurface in one (or possibly two) event(s). It is not possible to determine the timing of fracturing and subsequent collapse; however, the processes took place between the crater formation ($\ge$ 3.7 Ga) and late erosive events in Ares Vallis  (2.5 Ga). The order of magnitude of the total time span is therefore hundreds of million years within which sedimentation, melting of buried ice, fracturing and collapse occurred.
The amount of subsidence in Aram Chaos is directly related to the thickness of the buried sub-ice lake before mechanical destabilization. The fracturing was  a result of fast collapse of the sediment cap that occurred during the release of water toward the surface (Fig. 13a).

The formative time scale and water volume determination indicates the occurrence of very rapid (tens of days) and catastrophic (volume of flood of 9.3$\cdot$10$^4$ km$^3$) events capable of carving the Aram Valley and two other small channels. This means that a large amount of water would have to have been available at the source of the outflows.
In a context such as that observed in Aram Chaos, a temporary lake generated by a massive expulsion of liquid water from the subsurface of the chaos and confined within the crater rims could be a valid solution to explain the morphology of the area (subsidence and fracturing), the time scale and amount of water required to carve Aram Valley.

In the present-day atmospheric pressure conditions of Mars, the ponding of water for a long time is not feasible. Furthermore, the thermo-chronology of martian meteorites suggests that for most of the past 4 Ga, ambient near-surface temperatures on Mars were unlikely to have been much higher than the present cold state \citep{Shuster2005}. For these reasons, a rapid outflow of water followed directly by rapid valley formation is more consistent with the Hesperian condition. Slow and multi-stage outflow over millions of years require higher atmospheric pressure (and possible higher surface temperatures) for which no direct evidence exists. Under present-day conditions, lakes on Mars can remain largely liquid for thousands of years if an ice cap hinders the evaporation of water \citep{Newsom1996,Kreslavsky2002}, which would make the storage of large amounts of water possible for a relative short period and allow for the initial process of seepage erosion through the crater rim. In fact, given a typical groundwater flow velocity, seepage from the lake to Ares Vallis would have taken thousands of years.

\subsection{Early outflow event: two small channels}
After the collapse and rise of water, the first flow over the rim between Aram Chaos and Ares Vallis would have occurred locally and briefly at the topographically lowest point along the rim, which was localized in the NE sector at an elevation between -1800 and -2000 m (Figs 10 and 14a). Figure 13 shows that the -2000 m surface, excluding successive erosion, was likely mostly continuous around the crater, whereas the -1500 surface was localized in certain isolated areas. For this reason, an elevation between -1800 and -2000 m is a good approximation of the initial outflow channel inlet. Most initial erosion would have taken place at the downstream end where the flow descended steeply into Ares Vallis. The backward steps might have had certain antidunes in the initial stages, such as in the experiments of \citet{Kraal2008} and \citet{Marra2014} where the flow initially debouched onto a dry floor. The channel would have been narrower where it cut deeper into the plateau. On the upstream rim there would not have been much erosion because the flow essentially went uphill initially. With little or no sediment supply at the upstream, the channels slope quickly decreased from the initial values (0.028 -- 0.047, Figs 11c and green line of Fig. 14a) to the present-day slope (0.020, Fig. 10b). The lake water level decreased down to -2000 m and the outflow channels were abandoned (Fig. 14b).

\begin{figure}[ht!]
\centerline{\includegraphics[width=\columnwidth]{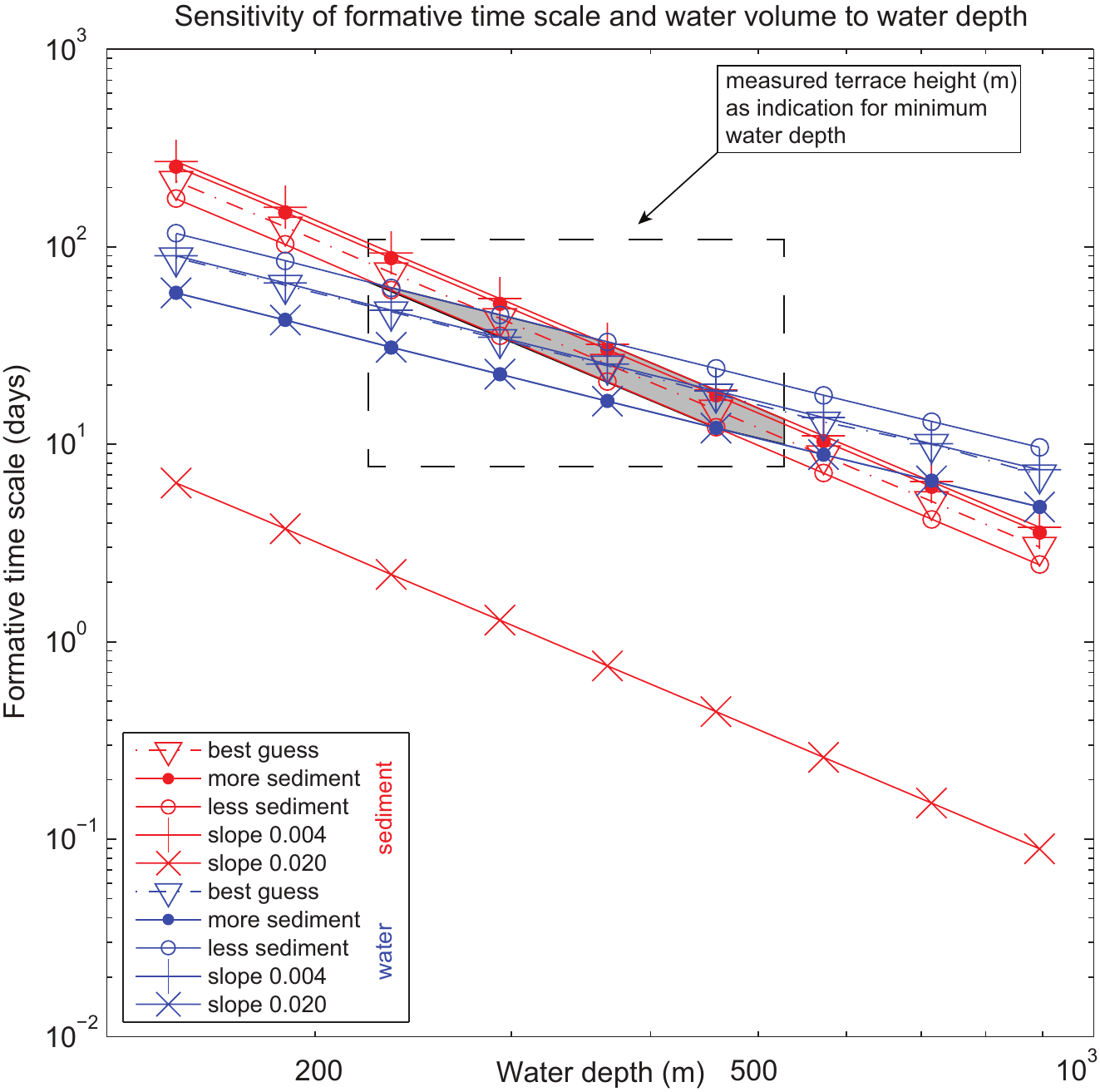}}
\caption{Sensitivity of formative time scale and water volume to water depth in the channel. The formative time scale for sediment ($T_s$, red lines) plotted versus the most uncertain parameter: water depth. Five lines are plotted for different combinations of gradient (slope) and larger or smaller sediment volume (Table 1). The formative time scale derived from the calculated flow discharge and measured volume of water ($T_w$, blue lines) in Aram Chaos is drawn for the same calculations. The range of depths for which $T_w$=$T_s$ matches the range of observed terrace depths (dashed square) and gives the likely time scale of tens of days (gray area).
}
\label{sensi}
\end{figure}

\subsection{Groundwater sapping as a precursor of the Aram Valley}
Temporary ponding of water would likely result in groundwater seepage and sapping along the eastern rim, where the subsurface gradient was steeper (Fig. 14a - blue lines). 
Therefore, together with the first outflow event, groundwater flow would have been present through the crater rim because of the relatively high porosity of the ejecta blanket. Where the groundwater emerged on the surface, certain erosional mechanisms became active because of the seepage \citep{Howard1988}. 
Three main processes of sediment mobilization, related to the seepage erosion, are described by \citet{Schorghofer2004} and \citet{Marra2014}: channelization, slumping, and fluidization.
Seepage erosion starts with a quick incision of an initial channel at the groundwater emergence point. This process is quite efficient because the critical mobilization slope of sediment is significantly smaller than the maximum angle of stability \citep{Lobkovsky2004,Marra2014}. 
The channel grows initially downwards and then upwards (headward). The channel profile is typically concave with an increase of slope toward the headcut. As a consequence, destabilization of the sediments surrounding the headcut area occurs and the slumping process starts. Continued headward erosion on the extending channel results in a migration of the groundwater flux toward the emergence point and activation of fluidification process. Therefore, the headward erosion rate increases rapidly through the crater rim. Finally, the breaching of rim occurs and the overland flow starts (Fig. 14b). 

\subsection{Late outflow event: Aram Valley final carving}
The initial overland flow conditions were characterized by a higher slope gradient (perhaps 0.02, as a remnant of the groundwater sapping, Fig 14b) and lower water depth (Table 1 - faster scenario). During the outflow event, the channel would have deepened, whereas the lake water level incrementally dropped as it emptied through the channel. The width of that channel was not necessarily equal to the width of the present valley, because the valley would have widened through slab failure and mass wasting while the channel undercut the valley sidewalls. The valley slope rapidly decreased from 0.02 to the present-day value of 0.004 (Fig. 14c, Table 1 - slower scenario).

In a waning flow, the sediment transport capacity would have reduced considerably so that the channel floor excavation no longer kept pace with the lake water level lowering. This would have ended the outflow event (Fig. 14c) and may have been abrupt or more gradual depending on the hypsometry of the crater basin immediately after the chaotic terrain collapse event. The formative time scale calculation (Table 1) and its sensitivity to water depth (Fig. 17) suggest that a rapid surface outflow event occurred and was active for tens of days.

The entire outflow process from Aram Chaos started shortly before and ended simultaneously with the late outflow event of Ares Vallis, although each single carving event lasted a maximum of tens of days. To have sufficient water volume available to carve the channels so rapidly, a lake at least 1500 m deep would have had to occur within Aram Chaos, which is consistent with the independent estimate for lake depth and subsidence.  

\begin{table*}
\centering
\caption{Comparison between formative models proposed for the chaotic terrains and Aram Chaos - Valley system: the water volume, number of floods, total time scale for carving, amount of subsidence and fractures occurrence suggested by the models are compared with those estimated for Aram Chaos.}
\normalsize
\begin{tabular}{lccccc}
\hline

Model&Water volume (km$^3$)&Events&Carving time scale (days)&subsidence (m)&Fracturing\\
\hline
Volcanic&10$^0$-10$^3$&$\ge$200&4000&$>$500&yes\\
Aquifer&10$^0$-10$^3$&$\ge$200&4000&max 500&pre-existing\\
Salt (gas) hydrated&10$^0$-10$^3$&few&4000&salt thickness&yes/pre-existing\\
Sub-ice lake&9.2$\cdot$10$^4$&1-2&20-40&1500&yes\\
&&&&&\\
Aram Chaos&9.3$\cdot$10$^4$&1-2& max 30&1500&yes\\

\hline
\label{}
\end{tabular}
\end{table*}

\section{Discussion}
In this section, we discuss certain open questions that remain regarding the buried sub-ice lake scenario and its possible applicability to the other chaotic terrains. We also present a possible connection between the proposed scenario and formation of the Interior Layered Deposits, which are frequently associated with chaotic terrains. 

\subsection{Open questions concerning the buried sub-ice lake scenario}
Although the sub-ice lake model is able to explain most characteristic features of chaotic terrains and outflow channels, certain questions remain open, such as those concerning the equatorial ice occurrence and long-term stability of the sub-ice lake. Although thick accumulations of ice on Mars occurred in high latitude areas, the location of Aram Chaos raises certain questions as to the process through which a several kilometers-thick ice deposit was initially formed in the Aram crater. During high-obliquity periods, snow/ice thickness that had accumulated in the equatorial regions could have reached centimeters over several years \citep{Jakosky1985,Mischna2003} rand resulted in a thick ice cover over millions of years. Many regions in the ancient southern highlands of Mars exhibited extensive fluvial activity, especially during the Noachian period. The sinks of the overland flow were craters and basins where alluvial fans, debris flow fans, deltas and sublacustrine fans are today observable on Mars \citep{Grotzinger2012}. Lakes generated from such craters and basins may have frozen and resulted in thick ice layers.  A thick ice deposit could also be accumulated by the freezing of an impact crater lake generated as a result of melting of the cryosphere induced by the increase in temperature related to the impact energy \citep{Newsom1996,Barnhart2010}.
A groundwater recharge from Ares Vallis during the earliest outflow event \citep[S1--S4,][]{Warner2009} and successive water freezing should also be considered, although the low subsurface gradient (Fig. 14c) makes this hypothesis unlikely. The remaining accessible observations in Aram Chaos do not allow for the choice of a most likely scenario for the formation of a thick ice fill of the initial crater, but the different options indicate that this is a feasible starting-point.

The sub-ice lake model implies the stability of a growing subsurface lake for several millions of years to achieve a liquid water layer thick enough to produce an approximately 10$^5$ km$^3$ outflow volume in one or possibly two events by using the geothermal heat flow as the sole heat source. This is only possible if the liquid water was contained in the subsurface and could not escape through the pore space and cracks in the surrounding solid rocks and ice. Under geologic timescales and at temperatures close to the melting point, water ice is extremely ductile and behaves like a viscous fluid, analogous to salt units in the subsurface on Earth \citep{Hudec2007}. The ice lid therefore acts as an impermeable lid to the liquid layer, closing any permeability that may form by cracks in the overburden.

The liquid water to the sides and bottom of the subsurface lake would be in direct contact with rocky material at a temperature above the melting curve \citep{Zegers2010}.Under particular conditions of porosity, permeability and hydraulic head, it would be possible for the sub-ice lake to drain and discharge through a subsurface aquifer away from Aram crater. However, such a scenario would produce a collapsed basin without a surface run-off feature, whereas in Aram Chaos, there is evidence of surface water run-off in the form of the Aram Valley.

\subsection{Buried sub-ice lake scenario and other chaotic terrains}
Other chaotic terrains on Mars share several characteristics with Aram Chaos, such as the fracturing and major subsidence of the fractured terrain, but differ in other respects. This has implications for the applicability of the buried sub-ice lake scenario for other chaotic terrains.
The buried sub-ice lake scenario needs the occurrence of a confined basin as initial condition, where an ice layer can accumulate and become buried by sediments afterwards. Some of the chaotic terrains on Mars have a clear circular shape suggesting an original impact crater (Aram Chaos, Marusky Chaos, Chaos SE of Hydaspis and SE of Pyrrhae, Orson Welles Chaos). The western part of Hydaspis Chaos shows circular fractures and round-shape boundaries that can be interpreted as the result of a collapse of two coalescent impact craters. 
Other chaotic terrains show irregular shapes with a random distribution of fractures and no circular crater shape is detectable. However, the occurrence of pre-existing basins can be deduced from the substantial subsidence characterizing all of the chaotic terrains. The irregular shape of the original basin may have been the result of coalescence of more than two impact craters, as can be observed in several regions of the Highlands. The occurrence of old fault bound basins cannot be excluded; however, it seems unlikely because the original shape would still be visible in the shape of chaotic terrains.

If certain basins occur, snow/ice thickness can be accumulated in the equatorial regions during high-obliquity periods or by partial melting of the cryosphere, producing liquid water to fill the basins before the subsequent lake freezing.
\citet{Hanna2007b} argue for extensive groundwater upwelling during the Late Noachian that was concentrated in the region around Xanthe Terra and Arabia Terra, where the water table exceeds the topography. The water can be achieved in pre-existing basins or craters during the Late Noachian. This can explain the concentration of the chaotic terrains around Xanthe Terra. The climate change occurring in the Early Hesperian would induce the lakes to freeze.

\citet{Zegers2010} show that planetary heat loss greater than 25 $mW/m^2$ is sufficient to melt 2 km thick ice layer if a sediment layer insulates ice from low surface temperatures. This is consistent with estimates of different authors and suggests the occurrence of planetary heat loss greater than 25 $mW/m^2$ during the Noachian-Hesperian period \citep{Hauck2002,Williams2004,Ruiz2011}, and it may justify the large-scale melting that occurred in the chaotic terrains region. However, a temporary increase in the surface and subsurface temperatures can be related to impacts \citep{Mangold2012} or magmatic intrusions \citep{Meresse2008}.

For Aram Chaos, the amount of water generated by the melting of buried ice layer was sufficient to carve the Aram Valley outflow channel. Although the calculation of water volume balance for other chaotic terrains and relative outflow channels is difficult because they were highly disturbed and modified by subsequent flows, larger outflow channels (e.g., Ares Vallis, Tiu Valles, Ravi Valles) seem to require larger amount of water than that achieved in the surrounding chaotic terrains. In these cases, perhaps a combination of the ice layer melting and other water supply processes occurred. However, a unique mechanism valid for all chaotic terrains has not been defined yet. Some mechanisms show particular features not shared with the other chaotic terrains. The question is if a unique mechanism is required to explain all of them or if each chaotic terrain can have one particular formative mechanism. Further studies and analysis are required to solve the controversy.

\subsection{Buried sub-ice lake scenario and Interior Layered Deposits}
The Interior Layered Deposits (ILDs) within Aram Chaos a have planar to gentle mound shape \citep{Glotch2005,Masse2008,Lichtenberg2010}. They were deposed after the chaotization event and are characterized, from bottom to top, by mono-hydrated sulfates, poly-hydrated sulfates and hematite layers. \citet{Rossi2008} suggest an Amazonian age for the ILDs of Aram Chaos. For their mineralogy, layering and shape, they are interpreted as the result of deposition in lacustrine or playa ambient \citep{Glotch2005,Arvidson2006} or as a result of multiple groundwater upwelling events \citep{Masse2008,Lichtenberg2010}. \citet{Rossi2008} argue for a possible spring origin.

Although the ILDs are clearly deposited after the chaotization process, the formation process of the chaotic terrain may be linked to the deposition of ILDs. In fact, the buried sub-ice scenario accounts for a number of pre-conditions that are favorable to the formation of sulfates: the formation of collapsed basins, where all ILDs occur,  and the presence of water to allow for the hydration of originally basaltic rocks.
The buried sub-ice lake scenario could also indirectly explain the origin of ILDs. Fractures occurring after the collapse may represent a preferential way for groundwater outflow, resulting in spring deposits as proposed by \citet{Rossi2008}. Perhaps multiple mechanisms acted at the same time or in sequence to form the chaotic terrains of Mars, but the evidence presented in this paper for the case of Aram Chaos demonstrates that chaotic terrain can form from catastrophic collapse of ice sheets.

\section{Conclusions}
The geological and hydrological analyses performed on the Aram Chaos-Valley system indicate that the flow volume required to carve the Aram Valley and two small channels (9.3$\cdot$10$^4$ km$^3$) is similar to the volume of water that could have been produced in an event of Aram Chaos by the melting of a buried ice lake \citep[9.2$\cdot$10$^4$ km$^3$,][]{Zegers2010}. The formative time evaluation confirms that a single, rapid (tens of days) and catastrophic event was sufficient to carve the channel rather than many small groundwater events active for a relatively long time. The resulting Aram Chaos morphology implies large amount of subsidence (1500 m) and one (or two) intense fracturing event(s). 

Such a scenario is consistent with a model of a buried sub-ice lake system \citep{Zegers2010}. The thermal insulation and relatively low heat flow may have been sufficient to induce ice melting. When the system became unstable, a massive water outflow occurred with the collapse of sediment and carving of the outflow channel. The sub-ice lake scenario explains many features characteristic of chaotic terrains, although more than one mechanism may have been involved in the carving of larger outflow channels.

\section{Acknowledgments}
Netherlands Organization for Scientific Research (NWO) and Netherlands Space Office (NSO grant) are gratefully acknowledged. We thank Wouter Marra for sharing insights in groundwater seepage erosion and catastrophic formation of valleys in experiments, Santiago de Beguera for stimulating discussion about debris-flow rheology and Rob Govers for useful comments. We gratefully acknowledge the anonymous reviewer and Timothy Glotch for their constructive criticism of the text. 
The authors contributed in the following proportions to conception and design, data collection, modeling, analysis and conclusions, and manuscript preparation (\%): MR(30,60,10,30,80); MGK(30,0,90,30,10); TEZ(30,0,0,30,10); JHPO(10,40,0,10,0).



\bibliographystyle{model2-names}
\bibliography{biblio}

\begin{thebibliography}{57}
\expandafter\ifx\csname natexlab\endcsname\relax\def\natexlab#1{#1}\fi
\expandafter\ifx\csname url\endcsname\relax
  \def\url#1{\texttt{#1}}\fi
\expandafter\ifx\csname urlprefix\endcsname\relax\def\urlprefix{URL }\fi
\providecommand{\eprint}[2][]{\url{#2}}
\providecommand{\bibinfo}[2]{#2}
\ifx\xfnm\relax \def\xfnm[#1]{\unskip,\space#1}\fi
\bibitem[{Andrews-Hanna and Phillips(2007)}]{Hanna2007a}
\bibinfo{author}{Andrews-Hanna, J.C.}, \bibinfo{author}{Phillips, R.J.},
  \bibinfo{year}{2007}.
\newblock \bibinfo{title}{{Hydrological modeling of outflow channels and chaos
  regions on Mars}}.
\newblock \bibinfo{journal}{Journal of Geophysical Research}
  \bibinfo{volume}{112}, \bibinfo{pages}{E08001}.
\bibitem[{Andrews-Hanna et~al.(2007)Andrews-Hanna, Phillips and
  Zuber}]{Hanna2007b}
\bibinfo{author}{Andrews-Hanna, J.C.}, \bibinfo{author}{Phillips, R.J.},
  \bibinfo{author}{Zuber, M.T.}, \bibinfo{year}{2007}.
\newblock \bibinfo{title}{{Meridiani Planum and the global hydrology of Mars}}.
\newblock \bibinfo{journal}{Nature} \bibinfo{volume}{446},
  \bibinfo{pages}{163--166}.
\bibitem[{Arvidson et~al.(2006)Arvidson, Poulet, Morris, Bibring, Bell,
  Squyres, Christensen, Bellucci, Gondet, Ehlmann, Farrand, Fergason, Golombek,
  Griffes, Grotzinger, Guinness, Herkenhoff, Johnson, Klingelhˆfer, Langevin,
  Ming, Seelos, Sullivan, Ward, Wiseman and Wolff}]{Arvidson2006}
\bibinfo{author}{Arvidson, R.E.}, \bibinfo{author}{Poulet, F.},
  \bibinfo{author}{Morris, R.V.}, \bibinfo{author}{Bibring, J.P.},
  \bibinfo{author}{Bell, J.~F., I.}, \bibinfo{author}{Squyres, S.W.},
  \bibinfo{author}{Christensen, P.R.}, \bibinfo{author}{Bellucci, G.},
  \bibinfo{author}{Gondet, B.}, \bibinfo{author}{Ehlmann, B.L.},
  \bibinfo{author}{Farrand, W.H.}, \bibinfo{author}{Fergason, R.L.},
  \bibinfo{author}{Golombek, M.}, \bibinfo{author}{Griffes, J.L.},
  \bibinfo{author}{Grotzinger, J.}, \bibinfo{author}{Guinness, E.A.},
  \bibinfo{author}{Herkenhoff, K.E.}, \bibinfo{author}{Johnson, J.R.},
  \bibinfo{author}{Klingelhˆfer, G.}, \bibinfo{author}{Langevin, Y.},
  \bibinfo{author}{Ming, D.}, \bibinfo{author}{Seelos, K.},
  \bibinfo{author}{Sullivan, R.J.}, \bibinfo{author}{Ward, J.G.},
  \bibinfo{author}{Wiseman, S.M.}, \bibinfo{author}{Wolff, M.},
  \bibinfo{year}{2006}.
\newblock \bibinfo{title}{{Nature and origin of the hematite-bearing plains of
  Terra Meridiani based on analyses of orbital and Mars Exploration rover data
  sets}}.
\newblock \bibinfo{journal}{Journal of Geophysical Research}
  \bibinfo{volume}{111}, \bibinfo{pages}{E1211}.
\bibitem[{Baker(2001)}]{Baker2001}
\bibinfo{author}{Baker, V.}, \bibinfo{year}{2001}.
\newblock \bibinfo{title}{{Water and the martian landscape}}.
\newblock \bibinfo{journal}{Nature} \bibinfo{volume}{412},
  \bibinfo{pages}{228--236}.
\bibitem[{Barnhart et~al.(2010)Barnhart, Nimmo and Travis}]{Barnhart2010}
\bibinfo{author}{Barnhart, C.J.}, \bibinfo{author}{Nimmo, F.},
  \bibinfo{author}{Travis, B.J.}, \bibinfo{year}{2010}.
\newblock \bibinfo{title}{{Martian post-impact hydrothermal systems
  incorporating freezing}}.
\newblock \bibinfo{journal}{Icarus} \bibinfo{volume}{208}, \bibinfo{pages}{101
  -- 117}.
\bibitem[{Brocard et~al.(2011)Brocard, Teyssier, Dunlap, Authemayou,
  Simon-Labric, Cacao-Chiqu{\'\i}n, Guti{\'e}rrez-Orrego and
  Mor{\'a}n-Ical}]{Brocard2011}
\bibinfo{author}{Brocard, G.}, \bibinfo{author}{Teyssier, C.},
  \bibinfo{author}{Dunlap, W.J.}, \bibinfo{author}{Authemayou, C.},
  \bibinfo{author}{Simon-Labric, T.}, \bibinfo{author}{Cacao-Chiqu{\'\i}n,
  E.N.}, \bibinfo{author}{Guti{\'e}rrez-Orrego, A.},
  \bibinfo{author}{Mor{\'a}n-Ical, S.}, \bibinfo{year}{2011}.
\newblock \bibinfo{title}{{Reorganization of a deeply incised drainage: role of
  deformation, sedimentation and groundwater flow}}.
\newblock \bibinfo{journal}{Basin Research} \bibinfo{volume}{23},
  \bibinfo{pages}{631--651}.
\bibitem[{Carr(1979)}]{Carr1979}
\bibinfo{author}{Carr, M.H.}, \bibinfo{year}{1979}.
\newblock \bibinfo{title}{{Formation of martian flood features by release of
  water from confined aquifers}}.
\newblock \bibinfo{journal}{Journal of Geophysical Research}
  \bibinfo{volume}{84}, \bibinfo{pages}{2995--3007}.
\bibitem[{Carr(1996)}]{Carr1996a}
\bibinfo{author}{Carr, M.H.}, \bibinfo{year}{1996}.
\newblock \bibinfo{title}{{Water on Mars}}.
\newblock \bibinfo{publisher}{Oxford University Press}.
\bibitem[{Chapman and Tanaka(2002)}]{Chapman2002}
\bibinfo{author}{Chapman, M.}, \bibinfo{author}{Tanaka, K.},
  \bibinfo{year}{2002}.
\newblock \bibinfo{title}{{Related magma-ice interactions: Possible origins of
  chasmata, chaos, and surface materials in Xanthe, Margaritifer, and Meridiani
  Terrae, Mars}}.
\newblock \bibinfo{journal}{Icarus} \bibinfo{volume}{155},
  \bibinfo{pages}{324--339}.
\bibitem[{Clifford(1993)}]{Clifford1993}
\bibinfo{author}{Clifford, S.M.}, \bibinfo{year}{1993}.
\newblock \bibinfo{title}{{A model for the hydrologic and climatic behavior of
  water on Mars}}.
\newblock \bibinfo{journal}{Journal of Geophysical Research}
  \bibinfo{volume}{98}, \bibinfo{pages}{10973--11016}.
\bibitem[{Clifford et~al.(2010)Clifford, Lasue, Heggy, Boisson, McGovern and
  Max}]{Clifford2010}
\bibinfo{author}{Clifford, S.M.}, \bibinfo{author}{Lasue, J.},
  \bibinfo{author}{Heggy, E.}, \bibinfo{author}{Boisson, J.},
  \bibinfo{author}{McGovern, P.}, \bibinfo{author}{Max, M.D.},
  \bibinfo{year}{2010}.
\newblock \bibinfo{title}{{Depth of the Martian cryosphere: Revised estimates
  and implications for the existence and detection of subpermafrost
  groundwater}}.
\newblock \bibinfo{journal}{Journal of Geophysical Research}
  \bibinfo{volume}{115}, \bibinfo{pages}{E07001}.
\bibitem[{Coleman(2005)}]{Coleman2005}
\bibinfo{author}{Coleman, N.M.}, \bibinfo{year}{2005}.
\newblock \bibinfo{title}{{Martian megaflood-triggered chaos formation,
  revealing groundwater depth, cryosphere thickness, and crustal heat flux}}.
\newblock \bibinfo{journal}{Journal of Geophysical Research}
  \bibinfo{volume}{110}, \bibinfo{pages}{E12S20}.
\bibitem[{Glotch and Christensen(2005)}]{Glotch2005}
\bibinfo{author}{Glotch, T.}, \bibinfo{author}{Christensen, P.},
  \bibinfo{year}{2005}.
\newblock \bibinfo{title}{{Geologic and mineralogic mapping of Aram Chaos:
  evidence for a water-rich history}}.
\newblock \bibinfo{journal}{Journal of Geophysical Research}
  \bibinfo{volume}{110}, \bibinfo{pages}{E09006}.
\bibitem[{Grotzinger and Milliken(2012)}]{Grotzinger2012}
\bibinfo{author}{Grotzinger, J.P.}, \bibinfo{author}{Milliken, R.E.},
  \bibinfo{year}{2012}.
\newblock \bibinfo{title}{{The Sedimentary Rock Record of Mars: Distribution,
  Origins, and Global Stratigraphy}}, in: \bibinfo{booktitle}{{Sedimentary
  Geology of Mars}}. \bibinfo{publisher}{SEPM (Society for Sedimentary
  Geology)}. volume \bibinfo{volume}{102}, pp. \bibinfo{pages}{1--48}.
\bibitem[{Harrison and Grimm(2008)}]{Harrison2008}
\bibinfo{author}{Harrison, K.P.}, \bibinfo{author}{Grimm, R.E.},
  \bibinfo{year}{2008}.
\newblock \bibinfo{title}{{Multiple flooding events in Martian outflow
  channels}}.
\newblock \bibinfo{journal}{Journal of Geophysical Research}
  \bibinfo{volume}{113}, \bibinfo{pages}{E02002}.
\bibitem[{Harrison and Grimm(2009)}]{Harrison2009}
\bibinfo{author}{Harrison, K.P.}, \bibinfo{author}{Grimm, R.E.},
  \bibinfo{year}{2009}.
\newblock \bibinfo{title}{{Regionally compartmented groundwater flow on Mars}}.
\newblock \bibinfo{journal}{Journal of Geophysical Research}
  \bibinfo{volume}{114}, \bibinfo{pages}{E04004}.
\bibitem[{Hauck and Phillips(2002)}]{Hauck2002}
\bibinfo{author}{Hauck, Steven~A., I.}, \bibinfo{author}{Phillips, R.J.},
  \bibinfo{year}{2002}.
\newblock \bibinfo{title}{{Thermal and crustal evolution of Mars}}.
\newblock \bibinfo{journal}{J. Geophys. Res.} \bibinfo{volume}{107},
  \bibinfo{pages}{5052}.
\bibitem[{Howard and McLane(1988)}]{Howard1988}
\bibinfo{author}{Howard, A.D.}, \bibinfo{author}{McLane, Charles~F., I.},
  \bibinfo{year}{1988}.
\newblock \bibinfo{title}{{Erosion of cohesionless sediment by groundwater
  seepage}}.
\newblock \bibinfo{journal}{Water Resour. Res.} \bibinfo{volume}{24},
  \bibinfo{pages}{1659--1674}.
\bibitem[{Hudec and Jackson(2007)}]{Hudec2007}
\bibinfo{author}{Hudec, M.R.}, \bibinfo{author}{Jackson, M.P.},
  \bibinfo{year}{2007}.
\newblock \bibinfo{title}{Terra infirma: Understanding salt tectonics}.
\newblock \bibinfo{journal}{Earth-Science Reviews} \bibinfo{volume}{82},
  \bibinfo{pages}{1 -- 28}.
\bibitem[{Hynek and Phillips(2001)}]{Hynek2001}
\bibinfo{author}{Hynek, B.}, \bibinfo{author}{Phillips, R.},
  \bibinfo{year}{2001}.
\newblock \bibinfo{title}{{Evidence for extensive denudation of the Martian
  highlands}}.
\newblock \bibinfo{journal}{Geology} \bibinfo{volume}{29},
  \bibinfo{pages}{407--410}.
\bibitem[{Jakosky and Carr(1985)}]{Jakosky1985}
\bibinfo{author}{Jakosky, B.M.}, \bibinfo{author}{Carr, M.H.},
  \bibinfo{year}{1985}.
\newblock \bibinfo{title}{{Possible precipitation of ice at low latitudes of
  Mars during periods of high obliquity}}.
\newblock \bibinfo{journal}{Nature} \bibinfo{volume}{315},
  \bibinfo{pages}{559--561}.
\bibitem[{van Kan~Parker et~al.(2010)van Kan~Parker, Zegers, Kneissl, Ivanov,
  Foing and Neukum}]{Parker2010}
\bibinfo{author}{van Kan~Parker, M.}, \bibinfo{author}{Zegers, T.},
  \bibinfo{author}{Kneissl, T.}, \bibinfo{author}{Ivanov, B.},
  \bibinfo{author}{Foing, B.}, \bibinfo{author}{Neukum, G.},
  \bibinfo{year}{2010}.
\newblock \bibinfo{title}{{3D structure of the Gusev Crater region}}.
\newblock \bibinfo{journal}{Earth and Planetary Science Letters}
  \bibinfo{volume}{294}, \bibinfo{pages}{411 -- 423}.
\bibitem[{Kargel et~al.(2007)Kargel, Furfaro, Prieto-Ballesteros, Rodriguez,
  Montgomery, Gillespie, Marion and Wood}]{Kargel2007}
\bibinfo{author}{Kargel, J.S.}, \bibinfo{author}{Furfaro, R.},
  \bibinfo{author}{Prieto-Ballesteros, O.}, \bibinfo{author}{Rodriguez,
  J.A.P.}, \bibinfo{author}{Montgomery, D.R.}, \bibinfo{author}{Gillespie,
  A.R.}, \bibinfo{author}{Marion, G.M.}, \bibinfo{author}{Wood, S.E.},
  \bibinfo{year}{2007}.
\newblock \bibinfo{title}{{Martian hydrogeology sustained by thermally
  insulating gas and salt hydrates}}.
\newblock \bibinfo{journal}{Geology} \bibinfo{volume}{35},
  \bibinfo{pages}{975--978}.
\bibitem[{Kleinhans(2005)}]{Kleinhans2005}
\bibinfo{author}{Kleinhans, M.G.}, \bibinfo{year}{2005}.
\newblock \bibinfo{title}{{Flow discharge and sediment transport models for
  estimating a minimum timescale of hydrological activity and channel and delta
  formation on Mars}}.
\newblock \bibinfo{journal}{Journal of Geophysical Research}
  \bibinfo{volume}{110}, \bibinfo{pages}{E12003}.
\bibitem[{Kraal et~al.(2008)Kraal, van Dijk, Postma and Kleinhans}]{Kraal2008}
\bibinfo{author}{Kraal, E.R.}, \bibinfo{author}{van Dijk, M.},
  \bibinfo{author}{Postma, G.}, \bibinfo{author}{Kleinhans, M.G.},
  \bibinfo{year}{2008}.
\newblock \bibinfo{title}{{Martian stepped-delta formation by rapid water
  release}}.
\newblock \bibinfo{journal}{Nature} \bibinfo{volume}{451},
  \bibinfo{pages}{973--976}.
\bibitem[{Kreslavsky and Head(2002)}]{Kreslavsky2002}
\bibinfo{author}{Kreslavsky, M.A.}, \bibinfo{author}{Head, J.W.},
  \bibinfo{year}{2002}.
\newblock \bibinfo{title}{{Fate of outflow channel effluents in the northern
  lowlands of Mars: The Vastitas Borealis Formation as a sublimation residue
  from frozen ponded bodies of water}}.
\newblock \bibinfo{journal}{J. Geophys. Res.} \bibinfo{volume}{107},
  \bibinfo{pages}{5121}.
\bibitem[{Leask et~al.(2006)Leask, Wilson and Mitchell}]{Leask2006b}
\bibinfo{author}{Leask, H.J.}, \bibinfo{author}{Wilson, L.},
  \bibinfo{author}{Mitchell, K.L.}, \bibinfo{year}{2006}.
\newblock \bibinfo{title}{{Formation of Aromatum Chaos, Mars: Morphological
  development as a result of volcano-ice interactions}}.
\newblock \bibinfo{journal}{J. Geophys. Res.} \bibinfo{volume}{111},
  \bibinfo{pages}{E08071}.
\bibitem[{Lichtenberg et~al.(2010)Lichtenberg, Arvidson, Morris, Murchie,
  Bishop, Fernandez~Remolar, Glotch, Dobrea, Mustard, Andrews-Hanna and
  Roach}]{Lichtenberg2010}
\bibinfo{author}{Lichtenberg, K.A.}, \bibinfo{author}{Arvidson, R.E.},
  \bibinfo{author}{Morris, R.V.}, \bibinfo{author}{Murchie, S.L.},
  \bibinfo{author}{Bishop, J.L.}, \bibinfo{author}{Fernandez~Remolar, D.},
  \bibinfo{author}{Glotch, T.D.}, \bibinfo{author}{Dobrea, E.N.},
  \bibinfo{author}{Mustard, J.F.}, \bibinfo{author}{Andrews-Hanna, J.},
  \bibinfo{author}{Roach, L.H.}, \bibinfo{year}{2010}.
\newblock \bibinfo{title}{{Stratigraphy of hydrated sulfates in the sedimentary
  deposits of Aram Chaos, Mars}}.
\newblock \bibinfo{journal}{Journal of Geophysical Research}
  \bibinfo{volume}{115}, \bibinfo{pages}{E00D17}.
\bibitem[{Lobkovsky et~al.(2004)Lobkovsky, Jensen, Kudrolli and
  Rothman}]{Lobkovsky2004}
\bibinfo{author}{Lobkovsky, A.E.}, \bibinfo{author}{Jensen, B.},
  \bibinfo{author}{Kudrolli, A.}, \bibinfo{author}{Rothman, D.H.},
  \bibinfo{year}{2004}.
\newblock \bibinfo{title}{Threshold phenomena in erosion driven by subsurface
  flow}.
\newblock \bibinfo{journal}{J. Geophys. Res.} \bibinfo{volume}{109},
  \bibinfo{pages}{F04010}.
\bibitem[{Mangold et~al.(2012)Mangold, Kite, Kleinhans, Newsom, Ansan, Hauber,
  Kraal, Quantin and Tanaka}]{Mangold2012}
\bibinfo{author}{Mangold, N.}, \bibinfo{author}{Kite, E.},
  \bibinfo{author}{Kleinhans, M.}, \bibinfo{author}{Newsom, H.},
  \bibinfo{author}{Ansan, V.}, \bibinfo{author}{Hauber, E.},
  \bibinfo{author}{Kraal, E.}, \bibinfo{author}{Quantin, C.},
  \bibinfo{author}{Tanaka, K.}, \bibinfo{year}{2012}.
\newblock \bibinfo{title}{{The origin and timing of fluvial activity at
  Eberswalde crater, Mars}}.
\newblock \bibinfo{journal}{Icarus} \bibinfo{volume}{220}, \bibinfo{pages}{530
  -- 551}.
\bibitem[{Manker and Johnson(1982)}]{Manker1982}
\bibinfo{author}{Manker, J.P.}, \bibinfo{author}{Johnson, A.P.},
  \bibinfo{year}{1982}.
\newblock \bibinfo{title}{{Simulation of martian chaotic terrain and outflow
  channels}}.
\newblock \bibinfo{journal}{Icarus} \bibinfo{volume}{51},
  \bibinfo{pages}{121--132}.
\bibitem[{Marra et~al.(2014)Marra, Braat, Baar and Kleinhans}]{Marra2014}
\bibinfo{author}{Marra, W.A.}, \bibinfo{author}{Braat, L.},
  \bibinfo{author}{Baar, A.W.}, \bibinfo{author}{Kleinhans, M.G.},
  \bibinfo{year}{2014}.
\newblock \bibinfo{title}{{Valley formation by groundwater seepage, pressurized
  groundwater outbursts and crater-lake overflow in flume experiments with
  implications for Mars}}.
\newblock \bibinfo{journal}{Icarus} \bibinfo{volume}{232},
  \bibinfo{pages}{97--117}.
\bibitem[{Masse et~al.(2008)Masse, Le~Mouelic, Bourgeois, Combe, Le~Deit,
  Sotin, Bibring, Gondet and Langevin}]{Masse2008}
\bibinfo{author}{Masse, M.}, \bibinfo{author}{Le~Mouelic, S.},
  \bibinfo{author}{Bourgeois, O.}, \bibinfo{author}{Combe, J.P.},
  \bibinfo{author}{Le~Deit, L.}, \bibinfo{author}{Sotin, C.},
  \bibinfo{author}{Bibring, J.P.}, \bibinfo{author}{Gondet, B.},
  \bibinfo{author}{Langevin, Y.}, \bibinfo{year}{2008}.
\newblock \bibinfo{title}{{Mineralogical composition, structure, morphology,
  and geological history of Aram Chaos crater fill on Mars derived from OMEGA
  Mars Express data}}.
\newblock \bibinfo{journal}{Journal of Geophysical Research}
  \bibinfo{volume}{113}, \bibinfo{pages}{E12006}.
\bibitem[{Max and Clifford(2001)}]{Max2001}
\bibinfo{author}{Max, M.}, \bibinfo{author}{Clifford, S.},
  \bibinfo{year}{2001}.
\newblock \bibinfo{title}{{Initiation of Martian outflow channels: Related to
  the dissociation of gas hydrate?}}
\newblock \bibinfo{journal}{Geophysical Research Letters} \bibinfo{volume}{28},
  \bibinfo{pages}{1787--1790}.
\bibitem[{Meresse et~al.(2008)Meresse, Costard, Mangold, Masson and
  Neukum}]{Meresse2008}
\bibinfo{author}{Meresse, S.}, \bibinfo{author}{Costard, F.},
  \bibinfo{author}{Mangold, N.}, \bibinfo{author}{Masson, P.},
  \bibinfo{author}{Neukum, G.}, \bibinfo{year}{2008}.
\newblock \bibinfo{title}{{Formation and evolution of the chaotic terrains by
  subsidence and magmatism: Hydraotes Chaos, Mars}}.
\newblock \bibinfo{journal}{Icarus} \bibinfo{volume}{194},
  \bibinfo{pages}{487--500}.
\bibitem[{Mischna et~al.(2003)Mischna, Richardson, Wilson and
  McCleese}]{Mischna2003}
\bibinfo{author}{Mischna, M.}, \bibinfo{author}{Richardson, M.},
  \bibinfo{author}{Wilson, R.}, \bibinfo{author}{McCleese, D.},
  \bibinfo{year}{2003}.
\newblock \bibinfo{title}{{On the orbital forcing of Martian water and CO2
  cycles: A general circulation model study with simplified volatile schemes}}.
\newblock \bibinfo{journal}{Journal of Geophysical Research}
  \bibinfo{volume}{108}, \bibinfo{pages}{5062}.
\bibitem[{Montgomery and Gillespie(2005)}]{Montgomery2005}
\bibinfo{author}{Montgomery, D.}, \bibinfo{author}{Gillespie, A.},
  \bibinfo{year}{2005}.
\newblock \bibinfo{title}{{Formation of Martian outflow channels by
  catastrophic dewatering of evaporite deposits}}.
\newblock \bibinfo{journal}{Geology} \bibinfo{volume}{33},
  \bibinfo{pages}{625--628}.
\bibitem[{Nelson and Greeley(1999)}]{Nelson1999}
\bibinfo{author}{Nelson, D.}, \bibinfo{author}{Greeley, R.},
  \bibinfo{year}{1999}.
\newblock \bibinfo{title}{{Geology of Xanthe Terra outflow channels and the
  Mars Pathfinder landing site}}.
\newblock \bibinfo{journal}{Journal of Geophysical Research-Planets}
  \bibinfo{volume}{104}, \bibinfo{pages}{8653--8669}.
\bibitem[{Newsom et~al.(1996)Newsom, Brittelle, Hibbitts, Crossey and
  Kudo}]{Newsom1996}
\bibinfo{author}{Newsom, H.}, \bibinfo{author}{Brittelle, G.},
  \bibinfo{author}{Hibbitts, C.}, \bibinfo{author}{Crossey, L.},
  \bibinfo{author}{Kudo, A.}, \bibinfo{year}{1996}.
\newblock \bibinfo{title}{{Impact crater lakes on Mars}}.
\newblock \bibinfo{journal}{Journal of Geophysical Research}
  \bibinfo{volume}{101}, \bibinfo{pages}{14951--14955}.
\bibitem[{Ogawa et~al.(2003)Ogawa, Yamagishi and Kurita}]{Ogawa2003}
\bibinfo{author}{Ogawa, Y.}, \bibinfo{author}{Yamagishi, Y.},
  \bibinfo{author}{Kurita, K.}, \bibinfo{year}{2003}.
\newblock \bibinfo{title}{{Evaluation of melting process of the permafrost on
  Mars: Its implication for surface features}}.
\newblock \bibinfo{journal}{Journal of Geophysical Research}
  \bibinfo{volume}{108}, \bibinfo{pages}{8046}.
\bibitem[{Oosthoek et~al.(2007)Oosthoek, Zegers, Rossi, Foing, Neukum and the
  HRSC Co-Investigator~Team}]{Oosthoek2007a}
\bibinfo{author}{Oosthoek, J.}, \bibinfo{author}{Zegers, T.},
  \bibinfo{author}{Rossi, A.}, \bibinfo{author}{Foing, B.},
  \bibinfo{author}{Neukum, G.}, \bibinfo{author}{the HRSC
  Co-Investigator~Team}, \bibinfo{year}{2007}.
\newblock \bibinfo{title}{{3D mapping of Aram Chaos: a record of fracturing and
  fluid activity}}, in: \bibinfo{booktitle}{Lunar and Planetary Science
  XXXVIII}, pp. \bibinfo{pages}{1--2}.
\bibitem[{Pacifici et~al.(2009)Pacifici, Komatsu and Pondrelli}]{Pacifici2009}
\bibinfo{author}{Pacifici, A.}, \bibinfo{author}{Komatsu, G.},
  \bibinfo{author}{Pondrelli, M.}, \bibinfo{year}{2009}.
\newblock \bibinfo{title}{{Geological evolution of Ares Vallis on Mars:
  Formation by multiple events of catastrophic flooding, glacial and
  periglacial processes}}.
\newblock \bibinfo{journal}{Icarus} \bibinfo{volume}{202}, \bibinfo{pages}{60
  -- 77}.
\bibitem[{Rodriguez et~al.(2005)Rodriguez, Sasaki, Kuzmin, Dohm, Tanaka,
  Miyamoto, Kurita, Komatsu, Fairen and Ferris}]{Rodriguez2005a}
\bibinfo{author}{Rodriguez, J.}, \bibinfo{author}{Sasaki, S.},
  \bibinfo{author}{Kuzmin, R.}, \bibinfo{author}{Dohm, J.},
  \bibinfo{author}{Tanaka, K.}, \bibinfo{author}{Miyamoto, H.},
  \bibinfo{author}{Kurita, K.}, \bibinfo{author}{Komatsu, G.},
  \bibinfo{author}{Fairen, A.}, \bibinfo{author}{Ferris, J.},
  \bibinfo{year}{2005}.
\newblock \bibinfo{title}{{Outflow channel sources, reactivation, and chaos
  formation, Xanthe Terra, Mars}}.
\newblock \bibinfo{journal}{Icarus} \bibinfo{volume}{175},
  \bibinfo{pages}{36--57}.
\bibitem[{Rossi et~al.(2008)Rossi, Neukum, Pondrelli, van Gasselt, Zegers,
  Hauber, Chicarro and Foing}]{Rossi2008}
\bibinfo{author}{Rossi, A.P.}, \bibinfo{author}{Neukum, G.},
  \bibinfo{author}{Pondrelli, M.}, \bibinfo{author}{van Gasselt, S.},
  \bibinfo{author}{Zegers, T.}, \bibinfo{author}{Hauber, E.},
  \bibinfo{author}{Chicarro, A.}, \bibinfo{author}{Foing, B.},
  \bibinfo{year}{2008}.
\newblock \bibinfo{title}{{Large-scale spring deposits on Mars?}}
\newblock \bibinfo{journal}{Journal of Geophysical Research}
  \bibinfo{volume}{113}, \bibinfo{pages}{E08016}.
\bibitem[{Ruiz et~al.(2011)Ruiz, McGovern, Jim{\'e}nez-D{\'\i}az, L{\'o}pez,
  Williams, Hahn and Tejero}]{Ruiz2011}
\bibinfo{author}{Ruiz, J.}, \bibinfo{author}{McGovern, P.J.},
  \bibinfo{author}{Jim{\'e}nez-D{\'\i}az, A.}, \bibinfo{author}{L{\'o}pez, V.},
  \bibinfo{author}{Williams, J.P.}, \bibinfo{author}{Hahn, B.C.},
  \bibinfo{author}{Tejero, R.}, \bibinfo{year}{2011}.
\newblock \bibinfo{title}{{The thermal evolution of Mars as constrained by
  paleo-heat flows}}.
\newblock \bibinfo{journal}{Icarus} \bibinfo{volume}{215}, \bibinfo{pages}{508
  -- 517}.
\bibitem[{Schorghofer et~al.(2004)Schorghofer, Jensen, Kudrolli and
  Rothman}]{Schorghofer2004}
\bibinfo{author}{Schorghofer, N.}, \bibinfo{author}{Jensen, B.},
  \bibinfo{author}{Kudrolli, A.}, \bibinfo{author}{Rothman, D.H.},
  \bibinfo{year}{2004}.
\newblock \bibinfo{title}{Spontaneous channelization in permeable ground:
  theory, experiment, and observation}.
\newblock \bibinfo{journal}{Journal of Fluid Mechanics} \bibinfo{volume}{503},
  \bibinfo{pages}{357--374}.
\bibitem[{Schultz et~al.(1982)Schultz, Schultz and Rogers}]{Schultz1982}
\bibinfo{author}{Schultz, P.H.}, \bibinfo{author}{Schultz, R.A.},
  \bibinfo{author}{Rogers, J.}, \bibinfo{year}{1982}.
\newblock \bibinfo{title}{{The Structure and Evolution of Ancient Impact Basins
  on Mars}}.
\newblock \bibinfo{journal}{Journal of Geophysical Research}
  \bibinfo{volume}{87}, \bibinfo{pages}{9803--9820}.
\bibitem[{Schumacher and Zegers(2011)}]{Schumacher2011}
\bibinfo{author}{Schumacher, S.}, \bibinfo{author}{Zegers, T.E.},
  \bibinfo{year}{2011}.
\newblock \bibinfo{title}{{Aram Chaos and its constraints on the surface heat
  flux of Mars}}.
\newblock \bibinfo{journal}{Icarus} \bibinfo{volume}{211}, \bibinfo{pages}{305
  -- 315}.
\bibitem[{Sharp(1973)}]{Sharp1973}
\bibinfo{author}{Sharp, R.P.}, \bibinfo{year}{1973}.
\newblock \bibinfo{title}{{Mars - Fretted and chaotic terrains}}.
\newblock \bibinfo{journal}{Journal of Geophysical Research}
  \bibinfo{volume}{78}, \bibinfo{pages}{4073--4083}.
\bibitem[{Shuster and Weiss(2005)}]{Shuster2005}
\bibinfo{author}{Shuster, D.}, \bibinfo{author}{Weiss, B.},
  \bibinfo{year}{2005}.
\newblock \bibinfo{title}{{Martian surface paleotemperatures from
  thermochronology of meterorites}}.
\newblock \bibinfo{journal}{Science} \bibinfo{volume}{309},
  \bibinfo{pages}{594--597}.
\bibitem[{Tanaka et~al.(2003)Tanaka, Skinner, Hare, Joyal and
  Wenker}]{Tanaka2003}
\bibinfo{author}{Tanaka, K.}, \bibinfo{author}{Skinner, J.},
  \bibinfo{author}{Hare, T.}, \bibinfo{author}{Joyal, T.},
  \bibinfo{author}{Wenker, A.}, \bibinfo{year}{2003}.
\newblock \bibinfo{title}{{Resurfacing history of the northern plains of Mars
  based on geologic mapping of Mars Global Surveyor data}}.
\newblock \bibinfo{journal}{Journal of Geophysical Research-Planets}
  \bibinfo{volume}{108}, \bibinfo{pages}{8043}.
\bibitem[{Warner et~al.(2010a)Warner, Gupta, Lin, Kim, Muller and
  Morley}]{Warner2010b}
\bibinfo{author}{Warner, N.}, \bibinfo{author}{Gupta, S.},
  \bibinfo{author}{Lin, S.Y.}, \bibinfo{author}{Kim, J.R.},
  \bibinfo{author}{Muller, J.P.}, \bibinfo{author}{Morley, J.},
  \bibinfo{year}{2010}a.
\newblock \bibinfo{title}{{Late Noachian to Hesperian climate change on Mars:
  Evidence of episodic warming from transient crater lakes near Ares Vallis}}.
\newblock \bibinfo{journal}{Journal of Geophysical Research}
  \bibinfo{volume}{115}, \bibinfo{pages}{E06013}.
\bibitem[{Warner et~al.(2009)Warner, Gupta, Muller, Kim and Lin}]{Warner2009}
\bibinfo{author}{Warner, N.}, \bibinfo{author}{Gupta, S.},
  \bibinfo{author}{Muller, J.P.}, \bibinfo{author}{Kim, J.R.},
  \bibinfo{author}{Lin, S.Y.}, \bibinfo{year}{2009}.
\newblock \bibinfo{title}{{A refined chronology of catastrophic outflow events
  in Ares Vallis, Mars}}.
\newblock \bibinfo{journal}{Earth and Planetary Science Letters}
  \bibinfo{volume}{288}, \bibinfo{pages}{58--69}.
\bibitem[{Warner et~al.(2010b)Warner, Gupta, Kim, Lin and Muller}]{Warner2010a}
\bibinfo{author}{Warner, N.H.}, \bibinfo{author}{Gupta, S.},
  \bibinfo{author}{Kim, J.R.}, \bibinfo{author}{Lin, S.Y.},
  \bibinfo{author}{Muller, J.P.}, \bibinfo{year}{2010}b.
\newblock \bibinfo{title}{{Retreat of a giant cataract in a long-lived
  (3.7--2.6 Ga) martian outflow channel}}.
\newblock \bibinfo{journal}{Geology} \bibinfo{volume}{38},
  \bibinfo{pages}{791--794}.
\bibitem[{Warner et~al.(2011)Warner, Gupta, Kim, Muller, Le~Corre, Morley, Lin
  and McGonigle}]{Warner2011}
\bibinfo{author}{Warner, N.H.}, \bibinfo{author}{Gupta, S.},
  \bibinfo{author}{Kim, J.R.}, \bibinfo{author}{Muller, J.P.},
  \bibinfo{author}{Le~Corre, L.}, \bibinfo{author}{Morley, J.},
  \bibinfo{author}{Lin, S.Y.}, \bibinfo{author}{McGonigle, C.},
  \bibinfo{year}{2011}.
\newblock \bibinfo{title}{{Constraints on the origin and evolution of Iani
  Chaos, Mars}}.
\newblock \bibinfo{journal}{Journal of Geophysical Research}
  \bibinfo{volume}{116}, \bibinfo{pages}{E06003}.
\bibitem[{Williams and Nimmo(2004)}]{Williams2004}
\bibinfo{author}{Williams, J.P.}, \bibinfo{author}{Nimmo, F.},
  \bibinfo{year}{2004}.
\newblock \bibinfo{title}{{Thermal evolution of the Martian core: Implications
  for an early dynamo}}.
\newblock \bibinfo{journal}{Geology} \bibinfo{volume}{32},
  \bibinfo{pages}{97--100}.
\bibitem[{Zegers et~al.(2010)Zegers, Oosthoek, Rossi, Blom and
  Schumacher}]{Zegers2010}
\bibinfo{author}{Zegers, T.E.}, \bibinfo{author}{Oosthoek, J.H.},
  \bibinfo{author}{Rossi, A.P.}, \bibinfo{author}{Blom, J.K.},
  \bibinfo{author}{Schumacher, S.}, \bibinfo{year}{2010}.
\newblock \bibinfo{title}{{Melt and collapse of buried water ice: An
  alternative hypothesis for the formation of chaotic terrains on Mars}}.
\newblock \bibinfo{journal}{Earth and Planetary Science Letters}
  \bibinfo{volume}{297}, \bibinfo{pages}{496 -- 504}.

\end{thebibliography}




\end{document}